 \documentclass[twocolumn]{aastex631}
\shorttitle{Multi-wavelength study of EHBL source 1ES\,0229$+$200}
\shortauthors{Hota et al.}
\usepackage{csquotes}
\usepackage{newtxtext,newtxmath}
\usepackage{hyperref}
\usepackage{booktabs}
\usepackage{lipsum} 
\usepackage{comment}
\usepackage{graphicx}	% Including figure files
\usepackage{amsmath}	% Advanced maths commands
\begin{document}

\title{Multi-wavelength study of extreme high-energy peaked BL Lac (EHBL) source 1ES\,0229$+$200 using ultraviolet, X-ray and gamma-ray observations}% using gamma-ray, X-ray, and ultra-violet observations}

\correspondingauthor{Jyotishree Hota}
\email{hotajyoti@gmail.com, acp.phy@gmail.com}

\author{Jyotishree Hota} %ORCID 0000-0002-0786-7307
\affiliation{Department of Physics and Astronomy, National Institute of Technology, Rourkela $-$ 769008, India\\}

\author{Rukaiya Khatoon }
\affiliation{Centre for Space Research, North-West University, Potchefstroom $-$ 2531, South Africa\\}

\author{Ranjeev Misra }
\affiliation{Inter-University Center for Astronomy and Astrophysics, Post Bag 4, Ganeshkhind, Pune $-$ 411007, India}

\author{Ananta C. Pradhan }
\affiliation{Department of Physics and Astronomy, National Institute of Technology, Rourkela $-$ 769008, India\\}

%\collaboration{20}{(AAS Journals Data Editors)}

%\author{F.X Timmes}
%\affiliation{Arizona State University}
%\affiliation{AAS Journals Associate Editor-in-Chief}

%\altaffiliation{AASTeX v6+ programmer}

%\affiliation{AAS Director of Publishing}
%\affiliation{American Astronomical Society \\1667 K Street NW, Suite 800 \\Washington, DC 20006, USA}



\begin{abstract}

We present a comprehensive analysis of the broadband spectral energy distribution (SED)  of the extreme high-energy peaked BL Lac (EHBL) source, 1ES 0229$+$200. Our study utilizes near-simultaneous data collected at various epochs between September 2017 and August 2021 (MJD: 58119$-$59365) from different instruments, including {\em AstroSat}$-$UVIT, SXT, LAXPC, {\em Swift}$-$UVOT, {\em Fermi}-LAT, and MAGIC. We investigate the one-zone synchrotron and synchrotron self-Compton (SSC) model, employing diverse particle distributions such as the log parabola, broken power law, power law with a maximum electron energy $\gamma$, energy-dependent diffusion (EDD), and energy-dependent acceleration (EDA) models to fit the broadband SED of the source. Our findings indicate that both peaks in the SED are well described by the one-zone SSC model across all particle distribution models. We estimate the jet power for different particle distributions. The estimated jet power for broken power law particle distributions is found to be on the order of $10^{47}$ ($10^{44}$) erg s$^{-1}$ for a minimum electron energy $\gamma_{min}$ $\sim$10 (10$^4$). However, for intrinsically curved particle energy distributions (e.g., log parabola, EDD, and EDA models), the estimated jet power is $\sim$10$^{44}$ erg s$^{-1}$. The SED fitting at five epochs enables us to explore the correlation between the derived spectral parameters of various particle distribution models. Notably, the observed correlations are inconsistent with the predictions in the power-law with a maximum $\gamma$ model, although the EDD and EDA models yield the correlations as expected. Moreover, the estimated physical parameter values are consistent with the model assumptions.

\end{abstract}

 %Classical Novae (251) --- Ultraviolet astronomy(1736) --- History of astronomy(1868) --- Interdisciplinary astronomy(804)


\section{Introduction}

Blazars are a sub-class of active galactic nuclei (AGN) whose relativistic jet points in the direction of the observer’s line of sight  \citep{Urry_1995, 2019Blandford}. The key characteristics of blazars are their non-thermal broadband spectra, strong radio and optical polarization, and fast variability \citep{2000Sambruna,Fan}. The spectral energy distribution (SED) of blazars shows a double bump that ranges from radio to very high energy (VHE) $\gamma$-ray \citep{1998Fossati}. According to the leptonic model, the first bump, which peaks in the optical to X-ray bands, is attributable to synchrotron emission while the second bump, which peaks in $\gamma$-ray energies, is explained by inverse Compton (IC) scattering of low-energy photons \citep{1982Urry,1985Ghisellini,1987Begelman,1995Blandford,1996Bloom,2004Sokolov}. 

The low-energy photons can be either synchrotron photons or photons that are external to the jet. When synchrotron photons serve as the target for IC scattering, it is called synchrotron self-Compton (SSC) \citep{1974Jones,1992Maraschi,1993Ghisellini}. On the other hand, the scattering of external photons is termed as the external-Compton process \citep{1992Dermer,1994Sikora,2000Bazejowski,2017Shah}. Furthermore, the high energy emission can be explained through hadronic processes such as the proton synchrotron process and pion production process \citep{1992Mannheim,2001Mucke}. 

%The low-energy photons can be either synchrotron photons, known as synchrotron elf-Compton (SSC) \citep{1974Jones,1992Maraschi,1993Ghisellini} or photons that are external to the jet termed as external Compton (EC) \citep{1992Dermer,1994Sikora,2000Bazejowski,2017Shah}. Furthermore, the high energy emission can be explained through hadronic processes such as the proton synchrotron process and pion production process \citep{1992Mannheim,2001Mucke}. 
 
Based on their spectral line in the optical bands, blazars are commonly classified into two types, such as flat spectrum radio quasars (FSRQs) and BL Lac objects. Further, depending upon their low-energy peak position, BL Lacs are categorized into three groups: low-energy peak (LBL), intermediate energy peak (IBL), and high-energy peak (HBL) \citep{1995ApJ...444..567P}.

In addition to the above classification, a new class of high-energy peak BL Lac sources, called extreme HBL (EHBL) sources, exhibit ambiguous spectral properties in high-energy emission. These sources show either a low-energy peak, a high-energy peak, or both, with peak frequencies surpassing $10^{17}$ Hz for the synchrotron peak and above one TeV for the IC peak \citep{Costamante,2020Biteau}.  As predicted by the blazar sequence \citep{2008Ghisellini,2017Ghisellini}, these objects are located at the upper edge of the peak frequency and are at the lowest luminosity end. %Predictably, these objects are expected to be extremely faint according to the blazar sequence. They are situated at the upper edge of the peak frequency and the lowest luminosity end.
%In addition to the above classification, a new class of high-frequency BL lac sources has been introduced, called extreme HBL (EHBL) sources, and they have ambiguous spectral properties in high energy emission. These sources have either a low/high energy peak or both the peaks located at very high frequencies \citep[the synchrotron peak frequency exceeding $10^{17}$ Hz and the IC peak above one TeV;][]{Costamante}. These objects are predicted to be extremely faint by the blazar sequence because they are located at the location of the upper edge of the peak frequency and the lowest luminosity end. 
Over the last decade, the exceptional operational results of the Imaging Atmospheric Cherenkov Telescopes such as H.E.S.S., MAGIC, and VERITAS have identified 14 sources at VHE (E $>$ 100 GeV) \citep{2019Foffano, 2019aMAGICCollaboration}. Among these sources, \citet{Costamante2018} and \citet{2019aMAGICCollaboration} have detected seven objects in TeV energies and categorized them as hard-TeV blazars (high-energy peak located above one TeV) and the remaining seven EHBL sources are categorized as soft-TeV blazars by \citet{Foffano2019}. %It is interesting to note 
Remarkably, it is noticed that two more sources, Mrk 501 and 1ES 1959$+$650, also show EHBL behaviour (hard-TeV spectra) during their flaring states \citep{Pian_1998, 2018MAGICCollaboration}.

The EHBL source, 1ES 0229$+$200 (RA = 38.202562$^{\circ}$, DEC: 20.28819$^{\circ}$), is located at a considerable distance with a redshift of $z = 0.14$, \citep{2005Woo,2007Aharonian,2009Tavecchio}). This blazar is known for emitting VHE gamma rays in the hard TeV range \citep{2014Aliu}. Initially discovered during the Einstein IPC Slew Survey \citep{1992Elvis}, it was later identified as a high-frequency peaked BL Lac source based on the location of its synchrotron peak \citep{2011Ackermann}.
H.E.S.S. detected the hard spectrum of this source in VHE emission, reaching up to 10 TeV, classifying it as a hard-TeV EHBL  \citep{2007A&AAharonian}. The source is also included in the third catalogue of {\em Fermi}-LAT, having been detected by {\em Fermi}-LAT after integrating four years of exposure time \citep{2015Acero, 2012Vovk}.
1ES 0229$+$200 stands out as an ideal object for studying the extragalactic background light \citep[EBL; ][]{2007A&AAharonian, 2010Kneiske} and the intergalactic magnetic field \citep[IGMF; ][]{2010Neronov, 2010Tavecchio, 2023Acciari} due to its distant and hard-spectrum nature as a blazar.

%The EHBL source, 1ES 0229$+$200 (RA = 38.202562 degree, DEC: 20.28819 degree) is quite distant (redshift $z = 0.14$, \citep{2005Woo,2007Aharonian,2009Tavecchio}) VHE $\gamma-$ray emitting hard-TeV blazar \citep{2014Aliu}. It was first discovered during the Einstein IPC Slew Survey \citep{1992Elvis} and identified as a high-frequency peaked BL Lac source according to the location of its synchrotron peak \citep{2011Ackermann}. The H.E.S.S. detected the hard spectrum of the source at VHE emission up to 10 TeV and classified it as hard-TeV EHBL \citep{2007A&AAharonian}. This source is mentioned in the third catalogue of Fermi-LAT as it was detected by Fermi-LAT later, after integrating the four years of exposure time of observation \citep{2015Acero, 2012Vovk}. 1ES 0229$+$200 is an ideal object for the study of extragalactic background light \citep[EBL; ][]{2007A&AAharonian, 2010Kneiske} and the intergalactic magnetic field \citep[IGMF; ][]{2010Neronov, 2010Tavecchio} as this is a distant hard-spectrum blazar.

Modelling of hard-TeV spectra of a source is a challenging task. Various one-zone leptonic SSC models have been developed to explain the VHE emission of the source 1ES 0229$+$200 \citep{Abdo_2011,2012Aleksi, 2014Aliu, 2014Tanaka, 2018Costamante, 2019Foffano, 2019Prandini, 2023Diwan}. Recently, a few more models have also been developed to explain the hard-TeV blazars. \citet{2021Zech} developed a new one-zone lepto-hadronic emission model for extreme-TeV blazars and showed that the acceleration in a single standing shock is capable of reproducing the jet emission. However, the re-acceleration on a second shock is required for the source like 1ES0229$+$200 (having hardest $\gamma-$ray spectra). \citet{2022Wei-Jian} investigated the one-zone hadronuclear (pp) model and found that the hard-TeV spectrum of 1ES0229$+$200 is reproduced %reduced
by the $\gamma-$ray emission %produced by 
from the $\pi^0$ decay in the p$-$p interactions. A lepto-hadronic model proposed by \citet{2022Aguilar-Ruiz} suggests that the two-zone emission can reproduce the broadband SEDs of the hard-TeV blazars. The two-zone lepto-hadronic model was also able to relax several parameters required by the one-zone models.

%In order to understand 
To understand the hard-TeV systems, one has to find the energy-generating mechanism responsible for the high jet power ($P_{jet}$) in such blazars. There are two mechanisms, Blandford–Znajek process \citep{1977Blandford} and Blandford–Payne process \citep{1982Blandford}, proposed for the jet formation in the hard-TeV blazars. The Blandford–Znajek model suggests that the blazar jet has an origin from a spinning black hole, and the $P_{jet}$ is associated with the spin and mass of the black hole with the magnetic field at its horizon, whereas the Blandford–Payne model suggests that the jet is originated from the black hole accretion disk. A recent study by \citet{2022Zhang} suggests that the relativistic jets are probably dominated by the Blandford–Znajek process for both FSRQs and BL Lacs. The $P_{jet}$ estimation in the one-zone SSC model is primarily governed by the electron energy distribution in the blob, which is characterized mostly by the broken power law with a minimum electron Lorentz factor ($\gamma_{min}$). \citet{2019Xue} and \citet{2021Zech} suggested a large $\gamma_{min}$ value ($\sim$$10^4$) is required to fit the SED in the TeV range spectra of the hard-TeV systems under one-zone SSC model. 
%Large minimum electron Lorentz factor values and rather low magnetization values appear necessary to fit the broadband SED for extreme-TeV blazars with typical one-zone emission models. %\citep{2021Zech}. 
\citet{2021Zech} %also 
proposed a standard one-zone lepto-hadronic emission model to study the broadband SED of TeV blazar %s,
1ES 0229$+$200, and they %have
calculated the total jet power $P_{jet}$ in the order of $\sim$$10^{44}$ ergs/sec. \citet{2020Acciari} applied the single-zone SSC model, spine-layer model, and Proton-Synchrotron Scenario (PSS) to fit the spectra of various extreme blazars and estimated their total jet powers. The jet power calculated using the spine–jet scenario ($10^{42}$ erg $sec^{-1}$) consistently appears more than one order of magnitude lower than those predicted by the SSC model ($10^{44}$ erg $sec^{-1}$). In contrast, the $P_{jet}$ estimated with the PSS is giving a higher value $\sim 0.15-45.6 \times 10^{46}$ erg $sec^{-1}$ and this is because PSS requires a rather large power in the protons responsible for the emission, often larger than the Eddington luminosity of the black hole powering the AGN \citep{2015Zdziarski}.

%\citet{2020Acciari} have fitted the spectra of several extreme blazars with the single-zone SSC model, spine$-$Layer model, and Proton$-$Synchrotron Scenario and estimated the total jet powers. The jet powers estimated with the spine–jet scenario ($10^{42}$ erg $sec^{-1}$) are systematically lower by more than one order of magnitude than those required by the SSC model ($10^{44}$ erg $sec^{-1}$). Whereas the jet powers estimated with Proton$-$Synchrotron Scenario is higher (0.15$-$45.6 $\times 10^{46}$ erg $sec^{-1}$ ). This is because they require a rather large power in the protons responsible for the emission, often larger than the Eddington luminosity of the black hole powering the AGN \citep{2015Zdziarski}. 

The $P_{jet}$ estimates mentioned above are primarily derived by assuming that the underlying electron energy distribution follows either a broken power law or a single power law with an exponential cutoff \citep{2020Acciari}. 
Nonetheless, there is evidence that suggests the underlying electron distribution, which may resemble a log-parabola distribution, has a considerable curvature in the observed spectrum \citep{2004Massaro, 2007Tramacere}.
%However, there is evidence suggesting that the observed spectrum exhibits significant curvature in the underlying electron distribution, possibly taking the form of a log-parabola distribution  \citep{2004Massaro, 2007Tramacere}.
%The above estimates of jet power are mainly based on assuming the underlying electron energy distribution is a broken power-law or a single power law with an exponential cutoff \citep{2020Acciari}. Some evidence suggests that the observed spectrum has a significant curvature in the underlying electron distribution which may have the shape of a log-parabola distribution \citep{2004Massaro, 2007Tramacere}. 
%However, there is evidence suggesting that the observed spectrum exhibits significant curvature in the underlying electron distribution, possibly taking the form of a log-parabola distribution (Massaro et al. 2004; Tramacere et al. 2007).
The curvature in the electron distribution is attributed to the energy dependency of particle acceleration/escape time-scales \citep{2017Sinha,2018Goswami,2021HOTA,2022Khatoon}. Additionally, physically motivated models have been employed, including distributions where the curvature of the spectrum can be attributed to the energy dependency of the diffusion time-scale (EDD) or due to the energy dependency of the acceleration time-scale (EDA), or with a high energy cutoff attributable to radiative losses ($\gamma$-max models) \citep{2021HOTA,2022Khatoon}.

%Additionally, physically motivated models have been employed, including distributions where the curvature of the spectrum can be attributed to the energy dependency of the diffusion time-scale (EDD) or due to the energy dependency of the acceleration time-scale (EDA), or with a high energy cutoff attributable to radiative losses (\xi-max models) (Hota et al. 2021; Khatoon et al. 2022)
%The energy dependency of the particle acceleration/escape time-scales is responsible for these curvatures in the electron distribution \citep{2017Sinha,2018Goswami,2021HOTA,2022Khatoon}. Moreover, physically motivated models were used, such as distributions where curvature of the spectrum can be attributed to the energy dependency of the diffusion time-scale (EDD) or due to energy dependency of the acceleration time-scale (EDA) or whose high energy cutoff is attributable to radiative losses ($\xi$-max models) \citep{2021HOTA,2022Khatoon}.

This study presents a standard SSC model for the broadband spectra of the hard-TeV blazar 1ES 0229$+$200. Various electron energy distribution models, such as the broken power-law model, log-parabola model, power-law particle distribution with maximum electron energy (PL with $\gamma_{max}$), energy-dependent diffusion (EDD), and energy-dependent acceleration (EDA) models, have been considered. These models have previously been verified for the X-ray spectral curvature of the HBL source, Mkn 421 \citep{2021HOTA, 2022Khatoon}.
Furthermore, a study on the broadband emissions of a HBL source Mkn501 has been conducted \citep{2024Bora}. % and the paper reporting this study is currently under review \citep[][under review]{2024hritwik}.
%and {\bf the broadband SED of the HBL source Mkn 501 \citep[][under review]{2024hritwik}}. 

%Here, we present a standard SSC model of the broadband spectra of a hard-TeV blazer 1ES 0229$+$200, considering various electron energy distribution models viz. broken power-law model, log-parabola model, power-law particle distribution with maximum electron energy, energy-dependent diffusion (EDD), and energy-dependent acceleration (EDA) models. These models are already verified for the X-ray spectral curvature of the HBL source, Mkn 421 \citep{2021HOTA, 2022Khatoon}. 

We have applied these particle distribution models to investigate the hard-TeV blazar source, extending the model for a broadband SED analysis using various observations over five epochs in a wide energy range, from ultraviolet (UV; 0.01 KeV) to VHE $\gamma-$ray ($\sim$10 TeV) bands.
%Here we have employed these particle distribution models to study the hard-TeV blazar source by extending the model for a broadband SED analysis using various observations over five epochs in a wide range of energy $-$ ultraviolet (UV; 0.01 KeV) to TeV $\gamma-$ray ($\sim$10 TeV) bands.

The paper is organized as follows: In Section \ref{sec:data_analysis}, we provide observations and data analysis procedure details. In Section \ref{sec:gammaray}, we present details of the $\gamma$-ray and VHE $\gamma$-ray analysis of the observed data. In Section \ref{sec:sedmodeling}, we discuss the broad-band SED modelling of the source with synchrotron and SSC emission processes. The results are presented in Subsection \ref{sec:jet} and \ref{sec:corel}. In Section \ref{sec:summary}, we summarized the results and their implications are discussed.

Throughout the paper, the following cosmological parameters are assumed$:$ $H_{0} = 70 km \ s^{-1} \ Mpc^{-1}$, $\Omega_{M}$= 0.3, $\Omega_{\Lambda}$= 0.7.

\section{Multi-wavelength observations and data analysis} \label{sec:data_analysis}

{\em AstroSat} is the first multi-wavelength space observatory of India, carrying five major payloads with energies ranging from UV to hard X-ray \citep{AGRAWAL20062989,KulinderPalSingh,2016arXiv160806051R}. {\em AstroSat} onboard instruments are Ultra-Violet Imaging Telescope \citep[UVIT: 130$-$300 nm; ][]{2017JApA...38...28T, 2017AJ....154..128T}, Soft X-ray focusing Telescope \citep[SXT: 0.3$-$8.0 keV; ][]{2017JApA...38...29S, 2016SPIE.9905E..1ES}, Large Area X-ray Proportional Counter \citep[LAXPC: 3$-$80 keV;][]{2016SPIE.9905E..1DY}, Cadmium Zinc Telluride Imager \citep[CZTI: 10$-$100 keV;][]{2017arXiv171010773R} and Scanning Sky Monitor (SSM). SXT was the primary instrument used to observe the blazar, 1ES0229$+$200 at five different epochs between 2017 and 2022. LAXPC and UVIT were also used to observe the source simultaneously. The {\em AstroSat} observation details of the source are listed in \autoref{tab:Astro_data}. Furthermore, for the LAXPC and SXT observations conducted during August 8–12, 2021, no simultaneous UVIT observations are available. Hence, we have used UV data from Swift/ultraviolet optical telescope (UVOT) observations for the above duration.

In this study, we have used the quasi-simultaneous observations %data 
from UVIT and UVOT for UV data, SXT, and LAXPC for X-ray data, {\em {Fermi}}-LAT for $\gamma$-ray data, and VHE spectral data from the MAGIC observations \citep{2019bMAGICCollaboration,MAGIC_Collaboration2020}. The details of the data reduction of UVIT, UVOT, SXT, LAXPC, and {\em Fermi}-LAT observations are as follows.

\begin{table*}
    \centering
     %\footnotesize
     \caption{Details of the {\em AstroSat} observations with LAXPC, SXT, and UVIT at five epochs.}
     \label{tab:Astro_data}
    %\resizebox{\columnwidth}{!}{
    \begin{tabular}{lcccccr}
    \hline
    \hline
    Observation ID & Instrument & Energy & Observation date &Exposure  & Count rate\\ %
    &&band & & (ks) & (Count/sec) \\
    \hline
        & LAXPC20 & 3-30 keV& 21-23, Sep 2017 & 57.4 & $2.74\pm 0.15 $\\
      A03$\_$078T01$\_$9000001546  &SXT & 0.7-7 keV & 21-23, Sep 2017 & 57.4 & $0.42\pm 0.01 $\\
      & UVIT & FUVBaF2 (1541 \AA) &22, Sep 2017 & 4.97   &$0.168 \pm 0.007$\\ 
     
      & UVIT& NUVB13(2447 \AA )&22, Sep 2017 & 5.0  &$0.463 \pm 0.01$\\
      \hline
     % & LAXPC 20 & 01-02,Oct 2017 \\
    %  A04$\_$130T01$\_$9000001572 & SXT & 01-02, Oct 2017   \\
   %   & UVIT & 01-Oct-2017  \\
   %   \hline
      & LAXPC20  & 3-30 keV & 9-10, Dec 2017 & 40 & $2.99\pm 0.10 $\\
      A04$\_$130T01$\_$9000001762 & SXT & 0.7-7 keV & 9-10, Dec 2017& 40 & $0.40\pm 0.02 $\\
      & UVIT & FUVBaF2 (1541 \AA) &9, Dec 2017& 4.97 &$0.175 \pm 0.006$\\
       & UVIT &  NUVB13(2447 \AA) &9, Dec 2017 & 4.97 &$0.468 \pm 0.01$\\
      \hline
      & LAXPC20  & 3-30 keV  & 21-22, Dec 2017 &50 & $3.15 \pm 0.10 $\\
      A04$\_$130T01$\_$9000001792 & SXT & 0.7-7 keV & 21-22, Dec 2017&50& $0.37 \pm 0.007 $\\
    & UVIT &  FUVBaF2 (1541 \AA) & 21, Dec 2017 & 4.99&$0.168 \pm 0.006$\\
    & UVIT &  NUVB13 (2447 \AA)  &21, Dec 2017 & 5.0  & $0.467 \pm 0.01$\\

      \hline
       & LAXPC20  & 3-30 keV  & 8-9, Dec 2018& 57.4 & $2.52\pm 0.06 $\\
      A04$\_$130T01$\_$9000001822  & SXT & 0.7-7 keV & 8-9, Jan 2018& 57.4 & $0.42 \pm 0.02 $\\
       & UVIT &  FUVBaF2(1541 \AA) & 8, Jan 2018 & 4.9 &$0.175 \pm 0.006$\\
       & UVIT   &  NUVB13 (2447 \AA ) &  8, Jan 2018& 5 & $0.50 \pm 0.01$\\
      \hline
    & LAXPC20  & 3-30 keV  & 8-12, Aug 2021& 343.6 & $2.32 \pm 0.09 $\\
    T04$\_$034T01$\_$9000004632 & SXT & 0.7-7 keV & 8-12, Aug 2021 &343.6 & $0.24 \pm 0.006 $\\
    & UVOT & W1(2600 \AA )& 8-12, Aug 2021& 3 & 0.613 $\pm$ 0.028 \\
     & UVOT & W2(1928 \AA) & 8-12, Aug 2021 &3 & 0.440 $\pm$ 0.021\\
     & UVOT & M2(2246 \AA) &8-12, Aug 2021&3  & 0.286 $\pm$ 0.015\\
\hline
    \end{tabular}
    %}
\end{table*}

\subsection{UVIT}

UVIT is an imaging telescope onboard {\em AstroSat} consisting of three channels: FUV (1300 - 1800 \AA), NUV (2000 - 3000 \AA), and Visible (3200 - 5500 \AA). The data reduction of the UVIT observations was performed with the software package CCDLAB \citep{2017Postma}. CCDLAB converts the Level 1 (L1) data into science-ready astronomical images. We extracted the L1 files of the UVIT observations and removed all the duplicate observations and short exposures (less than 120 seconds). The extracted L1 files were then supplied into CCDLAB for various corrections, viz., FPN, CPU distortion, flat-field, drift corrections, etc. The drift-corrected frames were then co-added to get the orbit-wise science-ready images for each observed filter. The images were co-aligned across all the orbits and merged into the final science images. Astrometric corrections were performed using the Gaia EDR3 catalogue. We applied the Source Extractor (SExtractor) tool \citep{1996Bertin} to extract the count rates in UVIT fits images corresponding to each filter. Then, the flux of the blazar source was calculated in all the UVIT filters using the unit conversion factor suggested by \citet{2017AJ....154..128T}. The flux values were then corrected for reddening using E(B$-$V) = 0.1192 values mentioned in \cite{Schlafly_2011}.

\subsection{UVOT}

The UVOT \citep[170-650 nm; ][]{2005Roming} onboard Swift uses three optical (U, B, V) and three UV (W1, M2, W2) filters to cover both the optical and UV portions of the spectrum. We have used Swift-UVOT observations to compliment the {\em AstroSat} SXT and LAXPC observations during 8$-$12 August, 2021 (at the fifth epoch) in the UV band. We have used observations in the W1, M2, and W2 filters of UVOT. We used the {\sc uvotimsum} % UVOTIMSUM
command to combine all the observed images in each UVOT filter. The task {\sc uvotsource} %UVOTSOURCE
was used to extract the magnitudes from the combined images. The source and background regions were selected with radii 5$''$ and 10$''$, respectively, as input parameters in the {\sc uvotsource} command. The observed magnitudes were then corrected for Galactic extinction using E(B$-$V) = 0.1192 mag and $R_{V} = 3.1$ \citep{Schlafly_2011}. We converted the UVOT magnitudes to flux units using the photometric zero-points from \citet{2011AIPC.1358..373B} and the conversion factors from \citet{2006Giommi}.

\subsection{SXT}

SXT is a focusing telescope with a field of view (FOV) of around $\sim$ 40$'$ diameter and operates in the soft X-ray energy range of 0.3 -- 8.0 keV for X-ray imaging and spectroscopy \citep{bf149366599c4681bceaadea3bdf114d}. The SXT observed 1ES 0229$+$200 in the Photon Counting (PC) mode for all the epoch and sxtpipeline\footnote{http://www.tifr.res.in/$^\sim$astrosat$\_$sxt/sxtpipeline.html} (AS1SXTLevel2, version 1.4b) was used to reduce the Level 1 data and obtain the cleaned Level 2 event files from different orbits. The standard Julia script developed by the instrument team was used to merge the cleaned event files from all orbits into a single file to avoid the problem of time-overlapping event files from successive orbits. The XSELECT (V2.4d) package built-in HEASOFT(V6.28) was used to extract the source spectrum from the processed Level-2 cleaned event files within a circular region of 15$'$ centred on the source. An off-axis auxiliary response file (ARF) was generated by using the SXT ARF generation tool\footnote{http://www.tifr.res.in/$^\sim$astrosat$\_$sxt/dataanalysis.html} with the help of the on-axis ARF (sxt$\_$pc$\_$excl00$\_$v04$\_$20190608.arf) as input. Further, we used ``SkyBkg$\_$comb$\_$EL3p5$\_$Cl$\_$Rd16p0$\_$v01.pha" for a background spectrum and ``sxt$\_$pc$\_$mat$\_$g0to12.rmf" for response matrix file  (RMF; given by the SXT team). The SXT spectrum was re-binned using the {\sc grppha} tool. For the spectral analysis, we used the energy range between 0.5$-$7.0 keV. To modify the gain of the response file, the gainfit tool in XSPEC was used with a fixed slope of value one and the offset as a free parameter. 
%\textbf{To consider Galactic absorption while fitting the broadband SED of the source, we used the TBabs model \citep{2000ApJ...542..914W} available in the XSPEC with the equivalent hydrogen column density ($N_{H}$) set at $7.9 \times 10^{20}$cm$^{-2}$  as determined by the online tool{\footnote{https://heasarc.gsfc.nasa.gov/cgi-bin/Tools/w3nh/w3nh.pl}} created by the LAB survey group \citep{2005A&A...440..775K}}
To account for the Galactic absorption, we have utilized the TBabs model \citep{2000ApJ...542..914W} available in the XSPEC for the spectral fit. The equivalent hydrogen column density ($N_{H}$) was fixed at $7.9 \times 10^{20}$cm$^{-2}$  for the X-ray observation, which was estimated from the online tool {\footnote{https://heasarc.gsfc.nasa.gov/cgi-bin/Tools/w3nh/w3nh.pl}}, by the LAB survey group \citep{2005A&A...440..775K}.

\subsection{LAXPC}

The LAXPC X-ray proportional counter has a high time resolution ($\sim$ 10 $\mu$s) and covers the energy range of 3$-$80 keV \citep{J.S.Yadav, 2017Antia, 2017Agrawal, Misra_2017}. It comprises three identical co-aligned proportional counter units, named LAXPC 10, LAXPC 20, and LAXPC 30, with the effective area of each detector as $\sim$2000 $cm^2$. All the Level 1 data were downloaded from the {\em AstroSat} archive. The Level 1 raw data were processed by the {\sc LAXPCSOFT} package (version as of 2022 August 15; which is recommended by the instrument team and accessible at the website of the {\em{AstroSat}} Science Support Cell\footnote{http://astrosat-ssc.iucaa.in}). LAXPC command laxpc$\_$make$\_$event were used to merge different orbits. To avoid the Earth occultation and the South Atlantic Anomaly, good time intervals (GTI files) were created using the command laxpc$\_$make$\_$stdgti. Finally, the source and background spectra were extracted by using the command laxpc$\_$make$\_$spectra. We have adopted the faint source method to extract the background spectrum and lightcurve \citep{2021arXiv210206402M}. Out of the three detectors, LAXPC 30 was switched off due to a gain instability issue arising from the gas leakage in the detector, and the LAXPC10 was working at low gain during the observation. Therefore, in our analysis, we have used only the LAXPC20 detector and limited the spectral analysis energy range to 4$-$18 keV.

\subsection{FERMI} \label{sec:fermi_obs}

The {\em Fermi} satellite carries the $\gamma$-ray instrument Large Area Telescope \citep[LAT;][]{2009Atwood}, which is sensitive in the energy range 20 MeV to 300 GeV. %It can scan the entire sky in about 3 hours and has a wide FOV $\sim$ 2.4 Sr. 
The {\em Fermi}-LAT continuously monitored more than 5000 extragalactic $\gamma$-ray sources \citep[4FGL;][]{Abdollahi_2020} between 2008 and 2018, out of which more than 3000 sources are blazars, suggesting that the $\gamma$-ray sky is heavily inhabited by relativistic jets. %Fermi-LAT has continuously observed the source 1ES 0229+200. 
In this analysis, we have collected the {\em Fermi}-LAT data of the source, 1ES 0229$+$200 from 2008$-$08$-$04 (MJD54682.6) to 2022$-$10$-$30 (MJD 59882). The analysis was carried out in the energy band 100 MeV$-$300 GeV using the latest version of fermipy$-$v0.17.4\footnote{Fermipy webpage: https://fermipy.readthedocs.io/en/latest/} and fermitools1$-$v1.2.23\footnote{ https://fermi.gsfc.nasa.gov/ssc/data/analysis/documentation/}. We selected a 15$^{\circ}$ region of interest (ROI) around the source to extract the photon events with evclass=128 and evtype=3, as recommended by the {\em Fermi}-LAT team in the fermitools documentation. The source model file was created using the Fermi 4FGL catalogue \citep{Abdollahi_2020}, and the background $\gamma$-ray emission and isotropic background emission were handled using  ``gll$\_$iem$\_$v07.fits'' and ``iso$\_$ P8R3$\_$ SOURCE$\_$ V3$\_$ v1.txt'' files, respectively. Additionally, a 90$^{\circ}$ zenith angle was chosen to prevent contamination from the earth limb. While extracting within the ROI, the source parameters were left thawed while those outside were frozen to their 4FGL catalogue values. The test statistics (TS) defined as TS = 2logL, where L is the likelihood parameter of the analysis \citep{1996Mattox} and is used to evaluate the detection significance of each source in the ROI.

\section{gamma-ray Analysis} \label{sec:gammaray}

%\begin{figure*}			
%\includegraphics[width=0.495\textwidth]{figures/fermi_magic_aN0_revised.pdf}\includegraphics[width=0.495\textwidth]{figures/fermi_magic_aN1_revised.pdf}\caption{Spectral fitting for combined $\gamma$-ray and VHE $\gamma$-ray data for 1ES 0229$+$200 using simple power law. }
%{\bf The left panel shows the combined spectral fit where aN value is set to zero for both Fermi-LAT and MAGIC datasets. The right panel shows the SPL fit where the aN value is set to one for the Fermi-LAT dataset, and the aN value is set to zero for the MAGIC dataset. We see the fitted residual for Fermi-LAT data is better when aN = 1 (right panel) compared to aN = 0 (left panel).} }\label{fig:sed_fermi_magic}\end{figure*}

\begin{figure}
  \centering
        \includegraphics[width=\columnwidth]{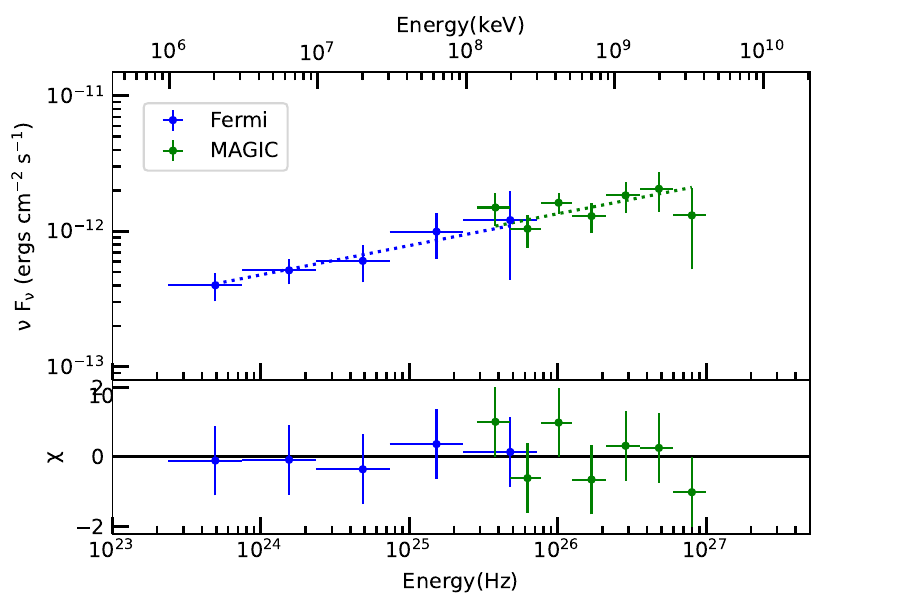}
   \caption{Spectral fitting with a simple power law model for the combined $\gamma$-ray (blue) and VHE $\gamma$-ray (green) data for 1ES 0229+200.}
   \label{fig:sed_fermi_magic}
\end{figure} 

This  Section presents the $\gamma$-ray spectral analysis of 1ES 0229$+$200 using quasi-simultaneous observations from {\em Fermi}-LAT and MAGIC. EHBLs are generally considered relatively faint sources in the high-energy $\gamma$-ray domain, owing to their low average brightness and the shifts in the IC peak location at higher energies.  The {\em Fermi}-LAT data considered in this work covers a substantial period from 2008 to 2022 ( Section \ref{sec:fermi_obs}). 

For the VHE $\gamma$-ray observations, we used the TeV spectra obtained from the MAGIC observations reported by \citet{MAGIC_Collaboration2020}, with a total exposure time of 117.46 hours and spanning a duration from 2013 to 2017. The VHE spectra has been corrected for extragalactic background light (EBL) absorption given by  \citet{2008Franceschini}.

 We undertake a joint fitting of the two, using a simple power law and the best fit was found to be with an index $\Gamma$ $\sim$1.78$\pm$0.3 and the fit resulted in a $\chi^2/dof = 4.28/10$. The best-fit spectra with residuals are shown in \autoref{fig:sed_fermi_magic}.
\label{eq:pow_conv}

\section{SED Modeling} \label{sec:sedmodeling}

\begin{figure*}
\centering
\includegraphics[width=0.9\columnwidth]{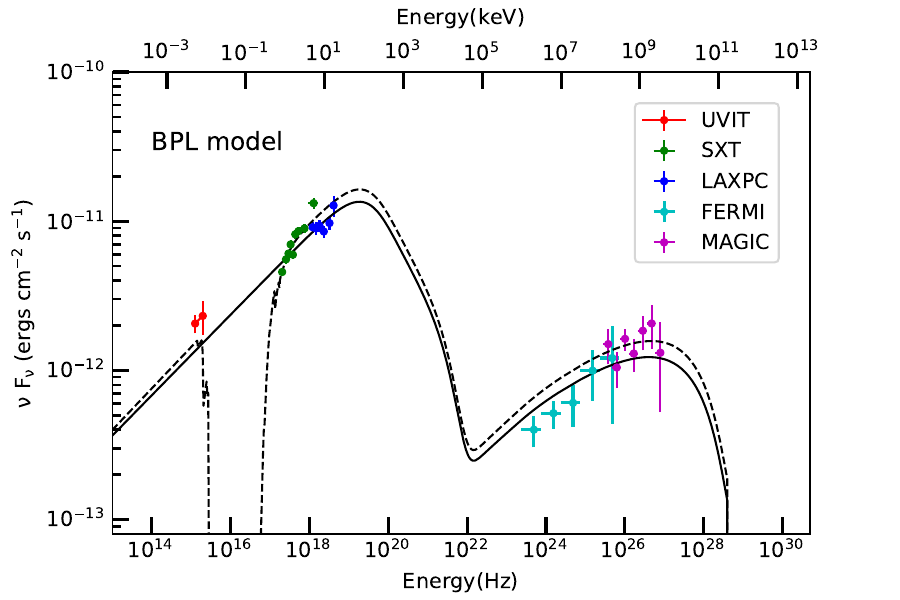}
\includegraphics[width=0.9\columnwidth]{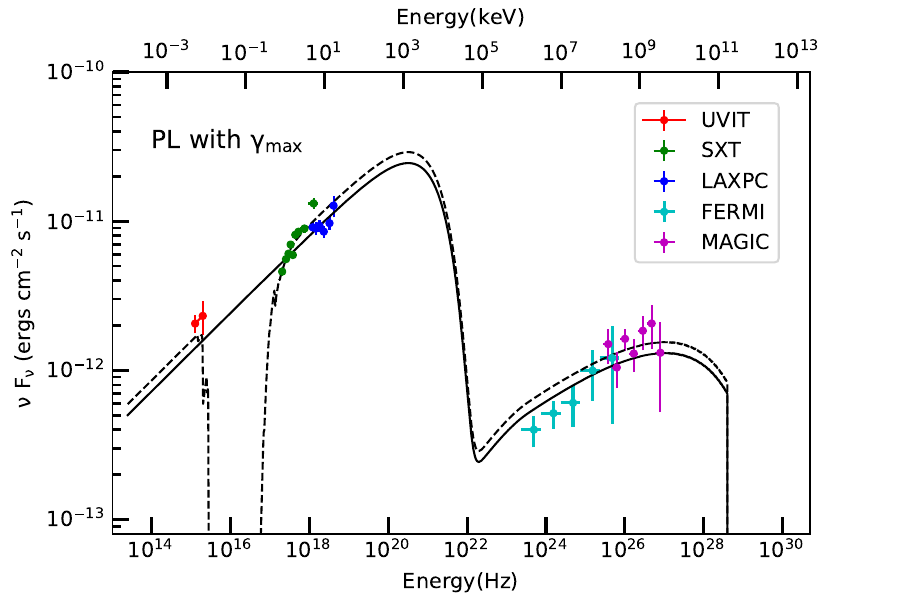}
\includegraphics[width=0.9\columnwidth]{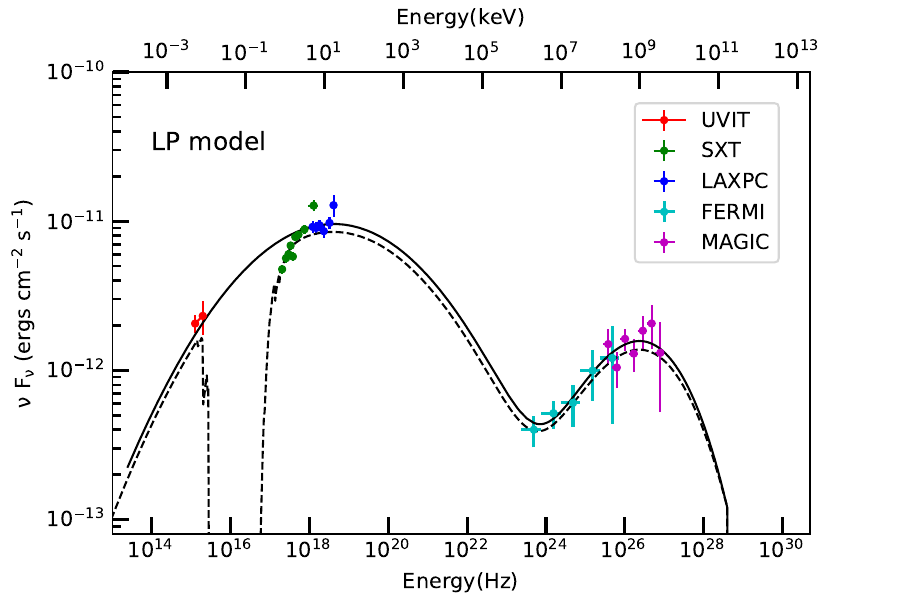}
\includegraphics[width=0.9\columnwidth]{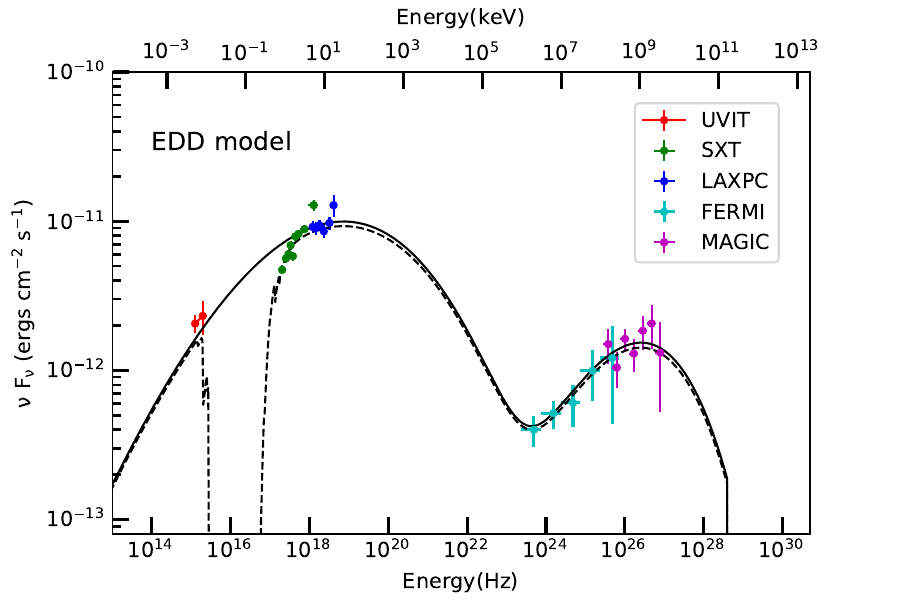}
\includegraphics[width=0.9\columnwidth]{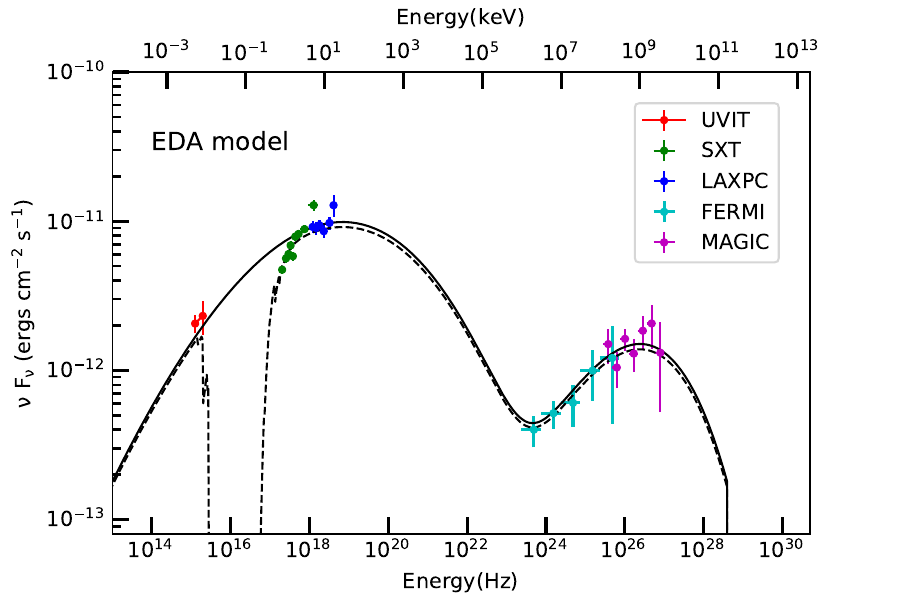}
\caption{Broadband spectral energy distribution (SED) of 1ES 0229$+$200 for data sets of epoch-4. Different panels show the particle distribution models (mentioned at the top left corner of each plot) used to construct the SEDs. The solid lines represent the best-fit model without Galactic absorption. The dotted lines represent the best-fit model spectra for the SXT data with Galactic absorption.}

%The solid line represents the outcome of the SSC model constructed using  UVIT, SXT, LAXPC and FERMI and MAGIC data.} %The dashed line is the outcome of the SSC model for MAGIC data.  The dashed-dotted line in the middle panel represents the SSC model output for SXT data. }%The dashed-dotted black line is the result of the conical jet SSC model. Details in the text.

\label{fig:sed_epoch2}
\end{figure*}

Broadband SED modelling of blazars is used to understand the underlying physical processes driving the broadband emission in both high-flux and low-flux states. 
%To carry out such a study, we needed simultaneous or quasi-simultaneous coverage of broad-band emissions. In this section, we study the broadband SED modelling of 1ES 0229$+$200 by using the data from UV to gamma-ray energy bands with Fermi-LAT, {\em AstroSat} $-$ LAXPC, SAXT, UVIT, and {\em Swift} $-$ SXT, UVOT observations. For very high energy data, we have used the high energy TeV spectral data points from the MAGIC observations \citep{MAGIC_Collaboration2020}. The VHE gamma-ray data are EBL corrected for the absorption effect.
We carried out broadband SED modelling of 1ES 0229$+$200 by using the data from UV to VHE $\gamma$-ray bands. %In the analysis of very high-energy spectra, we utilized the TeV spectral data from MAGIC observations as corrected for the EBL effect \citep{MAGIC_Collaboration2020}. The ftflx2xsp command converts the ASCII format of the $\gamma$-ray and TeV fluxes to pha format. {\em AstroSat} data can be considered quasi-simultaneous with  Fermi-LAT and MAGIC data. The fermi-LAT and MAGIC data are integrated over a long period due to the relatively faint emission. 
 % being integrated over a long period, due to the relatively faint emission.
 %The {\em AstroSat}$-$SXT/LAXPC/UVIT carried five observations in the low-activity state of 1ES 0229$+$200. 
%Details of the observations from various {\em {AstroSat}} instruments are provided in \autoref{tab:Astro_data}. Notably, no UVIT observations were available for the LAXPC and SXT observations conducted between 8$-$12 in August 2021. Consequently, {\em{ Swift}}$-$UVOT observations during this period were utilized in conjunction with AstroSat(SXT, LAXPC) data. 
The time intervals selected for the broadband spectral study were determined based on the availability of simultaneous observations in UV and X-ray energies. %We generated simultaneous multi-wavelength SEDs considering the observations at five epochs during the low-activity state of 1ES 0229$+$200. 
 We generated simultaneous multi-wavelength SEDs for all the five epochs of the observations of the source 1ES 0229$+$200 with {\em AstroSat}.
Details of the observations from {\em AstroSat} instruments are provided in \autoref{tab:Astro_data}.

%The details of the observations from various instrument of {\em AstroSat} is given in \autoref{tab:Astro_data}. No UVIT observations are available for the LAXPC and SXT observations carried out on 8$-$12, August 2021. % for the period during MJD: 59434 - 59438). Hence, we have used {\em Swift}$-$UVOT observation during this period of {\em AstroSat} (SXT, LAXPC) observations. % for broadband SED modeling of 1ES 0229+200. The time intervals for the broadband spectral study are chosen based on the availability of simultaneous observations in UV, X-ray and $\gamma$-ray energies. As a result, for the five-time periods (epochs) in the low-activity state of 1ES 0229+200, simultaneous multi-wavelength SEDs were generated. 

The one-zone SSC model provides an explanation for the broadband emission of EHBL sources like 1ES 0229$+$200. Leptonic models assume that interactions between relativistic electrons and the magnetic field in the emission zone produce the first hump of the SED in the energy range covering from radio to soft X-rays. Whereas the second hump of the SED is produced by IC scattering of a photon population, either the synchrotron photons themselves (SSC) and/or the photon field external to the jet (EC). Based on SSC models \citep{1993Ghisellini}, synchrotron photons generated by relativistic electrons in the magnetic field will be up-scattered. 
%Rfurther classified into synchrotron-self Compton (SSC) or external Compton (EC) categories based on the source of the seed photons.
We have employed a one-zone SSC leptonic model to explain the broadband SED fitting at each epoch. The model assumes a non-thermal emission from the blazar jet that arises from a spherical blob of radius R, filled with a tangled, homogeneous magnetic field (B), and isotropic electron distribution $n(\gamma)$ moving with the relativistic velocity along the jet. The blazer jet is moving with a bulk Lorentz factor, $\Gamma$, at an angle $\theta$ with respect to the observer's direction, affecting emission region by the beaming factor $\delta = 1/\Gamma(1 - \beta \cos{\theta})$. 

If the electron Lorentz factor, $\gamma$, is represented in terms of $\xi$ in such a way that $\xi=\gamma\sqrt{\mathbb{C}}$, where $\mathbb{C}=1.36 \times 10^{-11}\frac{\delta B}{1+z}$ with z being the redshift of source and $\delta$ is the jet Doppler factor, the synchrotron flux that the observer receives at energy $\epsilon$ will be \citep{Begelman},

%The synchrotron flux received by the observer at energy $\epsilon$ will be \citep{Begelman},
\begin{align}\label{eq:syn_conv}
	F_{\rm syn}(\epsilon) 
	%&= \frac{\delta^3(1+z)}{d_L^2} V  J_{\rm syn}\left(\frac{1+z}{\delta}\epsilon\right) \nonumber\\
	= \frac{\delta^3(1+z)}{d_L^2} V \,\mathbb{A} \int_{\xi_{min}}^{\xi_{max}} f(\epsilon/\xi^2)n(\xi)d\xi
\end{align}
here, V is the volume of the emission region, $d_L$ is the luminosity distance, $\mathbb{A}=\frac{\sqrt{3}\pi e^3B}{16m_ec^2\sqrt{\mathbb{C}}}$ and f(x) is the synchrotron emissivity function \citep{1986Rybicki}. 
 %\textbf{The electron lorentz factor, $\gamma$ is expressed in terms of $\xi$ such that $\xi=\gamma\sqrt{\mathbb{C}}$, where $\mathbb{C}=1.36 \times 10^{-11}\frac{\delta B}{1+z}$ with z being the redshift of source and $\delta$ is the jet Doppler factor. 
Rather than employing the electron's Lorentz factor, $\gamma$, the particle energy distribution is expressed by $n(\xi)$). Note that $\xi$ represents $\gamma$, and $\xi^2$ has dimension of keV. \citet{2021HOTA} solved \autoref{eq:syn_conv} numerically and included it as a local convolution model, \emph{$synconv \otimes n(\xi)$} in XSPEC \citep{1996Arnaud}. In this work, we have extended the numerical codes presented in \citet{2021HOTA} to estimate the emissivities and to calculate the observed fluxes corresponding to the synchrotron and SSC  emission processes. We have included it as a local convolution model, \emph{$sscicon \otimes n(\xi)$} in  XSPEC to fit the broadband SED of the source where $n(\xi)$ is the particle distribution.
%\textbf{To account for the Galactic absorption, we have utilized the TBabs model \citep{2000ApJ...542..914W} obtained from the online tool {\footnote{https://heasarc.gsfc.nasa.gov/cgi-bin/Tools/w3nh/w3nh.pl}} by the LAB survey group \citep{2005A&A...440..775K}. During the X-ray observations, the hydrogen column density ($N_{H}$) was kept fixed at 7.9 $\times$ $10^{20} cm^{−2}$.}
% Here, $\xi=\gamma\sqrt{\mathbb{C}}$, where $\mathbb{C}=1.36 \times 10^{-11}\frac{\delta B}{1+z}$ with z being the redshift of source and $\delta$ is the jet doppler factor.

%----------------------------------------- SED fit parameter table ---------------------------------------------------------------------------------------

\begin{table*}
\centering
%\footnotesize
\caption{The best-fitted spectral parameter values of the log-parabola, broken power law, PL with $\gamma_{max}$, EDD, and EDA models. The broadband SED fitting for the various models is constructed using observations from UV to gamma rays at each epoch. }
%{\bf $factor_{sxt}$ is an offset factor between LAXPC and SXT data.}}
\label{tab:sed_fitpar}
\vspace{0.3cm} 

\resizebox{\textwidth}{!}{ %

\begin{tabular}{lccccccr}
\hline \hline
S.N.  & Model & Unit & epoch-1 & epoch-2 & epoch-3 & epoch-4 & epoch-5 \\ %& epoch-6\\
 &  Parameters & & 21--23 Sep 17 & 9--10 Dec 17 & 21--22 Dec 17 & 8--9 Jan 18 & 8--12 Aug 21\\

\hline \hline
\multicolumn{8}{c}{log-parabola model ($constant*redden*TBabs*eblcor*sscicon*log-parabola$)} \\
\hline \hline

1 & $\alpha$ & - & 
$2.40_{-0.02}^{+0.02}$ & 
$2.65_{-0.02}^{+0.02}$&
$2.59_{-0.02}^{+0.02}$ & 
$2.67_{-0.02}^{+0.02}$ & 
$2.44_{-0.02}^{+0.02}$\\ \\
2 & $\beta$ & - & 
$0.34_{-0.01}^{+0.01} $ 
& $0.27_{-0.01}^{+0.01}$  
& $0.29_{-0.01}^{+0.01}$
&  $0.26_{-0.01}^{+0.01}$ 
& $0.32_{-0.01}^{+0.01}$\\ \\

3 & $N$ & ($10^{-11}$) 
& $2.39_{-0.2}^{+0.2}$ 
& $1.50_{-0.2}^{+0.2}$ 
&  $1.63_{-0.1}^{+0.1}$ 
& $1.42_{-0.2}^{+0.2}$ 
& $1.99_{-0.2}^{+0.2}$\\ \\
% 4  & $R$ & log(cm) &  17  & 17 & 17 & 17.0185 & 17\\
% 5 & $\Gamma$ & - & 20 & 20& 20 & 20& 20\\
4 & $B$ & ($10^{-3}$ G) 
& $1.02_{-0.1}^{+0.1}$ 
& $2.93_{-0.3}^{+0.3}$ 
&  $2.8_{-0.3}^{+0.3}$ 
& $3.04_{-0.5}^{+0.5}$&
$1.38_{-0.1}^{+0.1}$\\ \\
5 & $log \ P_{jet}$ &  & 
$43.9984_{-0.1}^{+0.1}$ &
$43.7553_{-0.1}^{+0.1}$ & 
$43.6735_{-0.1}^{+0.1}$ & 
$43.7544 _{-0.1}^{+0.1}$ & 
$43.8437_{-0.1}^{+0.1}$\\ \\
%5 & $ P_{jet}$ & $10^{44}$ erg/$sec$ &$0.89_{-0.20}^{+0.20}$ &$0.58_{-0.13}^{+0.13}$ & $0.5357_{-0.12}^{+0.12}$ &$0.54 _{-0.12}^{+0.12}$ & $0.892_{-0.20}^{+0.20}$\\ \\
6 & $\chi^{2}(dof)$ & - 
& 407.2(240) 
& 86.5(88)
& 138.8(107)
& 105(106) 
& 151.3(94)\\ \\
%7 & $factor_{magic}$ & - & $1.3_{-0.1}^{+0.1}$ & $1.6_{-0.4}^{+0.4}$ & $1.5_{-0.3}^{+0.3}$ & $1.7_{-0.6}^{+0.6}$& $1.4_{-0.2}^{+0.2}$\\ \\
%10& $factor_{magic}$ & - & & 0.84 & 0.74 & 0.69 & 0.87& 0.62\\
7 & $factor_{sxt}$ & - & 
$0.83_{-0.04}^{+0.04}$ & 
$0.76_{-0.02}^{+0.02}$& 
$0.75_{-0.02}^{+0.02}$ & 
$0.88_{-0.04}^{+0.04}$ & 
$0.63_{-0.03}^{+0.03}$\\\\
\hline \hline
\multicolumn{8}{c}{Broken power law model ($constant*redden*TBabs*eblcor*sscicon*bknpo$)} \\
\hline \hline

%1 & $\xi_{min}$ & ($10^{-6}$ $\sqrt{keV}$) &  6.04 & 9.97 & 10.0 & 11.4 & 6.18\\\\
 1 & $\gamma_{min}$ &   & 10 & 10 & 10 & 10 & 10\\\\
%2 & $\xi_{max}$ & $(\sqrt{keV})$ &   60.4 & 99.7 & 100 & 114 & 61.8\\\\
2 & $\gamma_{max}$ &  &   $10^8$ & $10^8$ & $10^8$ & $10^8$ & $10^8$ \\
\\
3 & $\xi_{break}$ & $\sqrt{keV}$ &  $> 10.5$ & $> 6.3$  &  $> 10.5$ & $ >7.8$ & $> 5.4$\\\\
%3 & $\gamma_{break}$ & ($10^6)$ &  $> 17.4$ & $> 6.32$  &  $> 10.5$ & $> 6.85 $& $> 8.74$\\\\
4 & $  p$ & - & 
$2.25_{-0.03}^{+0.03}$ &
$2.46_{-0.03}^{+0.03}$ 
& $2.40_{-0.03}^{+0.03}$ &
$2.55_{-0.03}^{+0.03}$ & 
$2.32_{-0.01}^{+0.01}$\\\\
5 & $q$ & - & 4.0  & 4.0 & 4.0 & 4.0& 4.0\\\\
6 & $N$ & ($10^{-12}$) &
$11.5_{-0.8}^{+0.8}$ & 
$09.10_{-0.6}^{+0.6}$ & 
$10.02_{-0.7}^{+0.7}$ & 
$8.68_{-0.7}^{+0.7}$ &
$10.7_{-0.9}^{+0.9}$\\\\
%7 & $R$ & log(cm) & 17 & 17& 17 & 17& 17\\\\
%8 & $\Gamma$ & - & 20& 20 & 20& 20 & 20\\\\
7 & $B$ &  ($10^{-3}$ G)& 
$1.67_{-0.3}^{+0.3}$  & 
$3.12_{-0.3}^{+0.3}$  &
$2.58_{-0.3}^{+0.3}$ &
$3.40_{-0.3}^{+0.3}$ &
$1.97_{-0.3}^{+0.3}$\\\\
8 & $log \ P_{jet}$ &  
&$46.0214_{0.06}^{0.06}$ 
& $45.4643_{0.01}^{0.01}$ 
& $46.5954_{0.01}^{0.01} $ 
&$46.69_{0.01}^{0.01}$
&$46.21_{0.008}^{0.008}$\\\\
%8 & $ P_{jet}$ & $10^{44}$erg/$sec$ &  $112.3_{-15.5}^{+15.5}$ & $31.59_{-0.73}^{+0.73}$ & $466.1_{-10.7}^{+10.7} $ & $972.07_{-22.4}^{+22.4}$ &$162.18_{-2.99}^{+2.99}$\\\\
9 & $\chi^{2}(dof)$ & - & 
310(243)& 
84.8(89) & 
110.2(108) & 
102.3(107)& 
112.4(95)\\\\
%10 & $factor_{magic}$ & - & $1.5_{-0.3}^{+0.3}$ & $3.8_{-0.6}^{+0.6}$& $2.5_{-0.5}^{+0.5}$ & $3.2_{-0.8}^{+0.8}$ & $1.8_{-0.4}^{+0.4}$\\\\
10 & $factor_{sxt}$ & - & 
$1.1_{-0.06}^{+0.06}$ & 
$0.98_{-0.06}^{+0.06}$& 
$0.90_{-0.06}^{+0.06}$ & 
$1.2_{-0.09}^{+0.09}$ & 
$0.81_{-0.04}^{+0.04}$\\\\

\hline 
\end{tabular} 
}
\end{table*}

\addtocounter{table}{-1}
\begin{table*}
%\ContinuedFloat
\centering
%\footnotesize
\caption{(Continued...) The best-fitted spectral parameter values of the log-parabola, broken power law, PL with $\gamma_{max}$, EDD, and EDA models. The broadband SED fitting for the various models is constructed using observations from UV to gamma rays at each epoch.}
\label{tab:sed_fitpar_cont}
\vspace{0.3cm} 
\resizebox{\textwidth}{!}{ %

\begin{tabular}{lccccccr}

\hline \hline
S.N.  & Model & Unit & epoch-1 & epoch-2 & epoch-3 & epoch-4 & epoch-5 \\ %& epoch-6\\
 & Parameters & & 21--23 sep 17 & 9--10 dec 17 & 21--22 dec 17 & 8--9 jan 18 & 8--12 aug 21\\
\hline \hline

%\hline \hline
\multicolumn{8}{c}{ PL with $\gamma_{max}$ model ($constant*redden*TBabs*eblcor*sscicon*\gamma_{max}$)} \\
\hline \hline

1 & $p$ & - &
$2.25_{-0.03}^{+0.03}$ & 
$2.51_{-0.05}^{+0.01}$ &
$2.42_{-0.09}^{+0.02}$ & 
$2.53_{-0.05}^{+0.01}$& 
$2.30_{-0.03}^{+0.03}$\\
2 & $\xi_{max}$ & - & $> 23 $& $> 17$ & $> 31.5$ & $> 8.8 $ & $> 13 $
\\
\\
3 & $N$ & $10^{-12}$ & 
$11.6_{-0.9}^{+0.9}$ &
$09.01_{-0.7}^{+0.24}$& 
$9.40_{-0.7}^{+0.7}$ & 
$8.60_{-0.6}^{+0.6}$&
$11.3_{-0.8}^{+0.8}$\\
\\
 %4 & $R$ & log(cm) & 17 & 17 & 17& 17 & 17\\\\
 %5 & $\Gamma$ & - & 20& 20 & 20& 20 & 20\\\\
4 & $B$ & $10^{-3}$G &
$1.56_{-0.4}^{+0.4}$ &
$4.14_{-0.5}^{+0.6}$& 
$3.68_{-0.5}^{+0.6}$ & 
$4.23_{-0.5}^{+0.5}$& 
$1.70_{-0.2}^{+0.2}$ \\
\\
5 & $log \ P_{jet}$ &  & 
$47.30_{-0.062}^{+0.062}$ &
$45.46_{0.067}^{ 0.067}$ & 
$48.07_{0.069}^{0.069}$ & 
$45.49_{0.07}^{0.07}$ & 
$45.15_{0.07}^{0.07}$\\
% 5 & $ P_{jet}$ & $10^{44}$ erg/$sec$ & $2085.4_{-297.7}^{+297.7}$ &  $31.45_{-4.85}^{+4.85}$ &  $10546.2_{-1675.6}^{+1675.6}$ & $34.03_{-5.48}^{+5.48}$ & $13.76_{-2.22}^{+2.22}$\\
 \\
6 & $\chi^{2}(dof)$ & - &
307.8(243) & 
82.5(89)&
108.2(108) & 
101.2(107)&
111.8(96)\\
\\
%8 & $factor_{magic}$ & - & $1.5_{-0.3}^{+0.3}$ & $3.7_{-0.3}^{+0.3}$& $2.2_{-0.3}^{+0.3}$ & $3.9 _{-0.3}^{+0.3}$& $2.1_{-0.3}^{+0.3}$\\
7 & $factor_{sxt}$ & - & $1.1_{-0.05}^{+0.05}$ & $1.01_{-0.05}^{+0.05}$ & $0.93_{-0.05}^{+0.05}$  & $1.1_{-0.05}^{+0.05}$  & $0.83_{-0.05}^{+0.05}$ \\
\\

\hline \hline
\multicolumn{8}{c}{EDD model ($constant*redden*TBabs*eblcor*sscicon*edd$)} \\
\hline \hline

1 & $B$ & $10^{-3}$ G
 & $1.26_{-0.1}^{+0.1}$
 & $1.44_{-0.3}^{+0.3}$ 
 %& $2.33_{-0.8}^{+0.8}$ 
 & $2.80_{-0.3}^{+0.3}$
 & $3.05_{-0.5}^{+0.5}$
 & $1.27_{-0.1}^{+0.1}$ \\ 
 \\
% 2 & $R$ & log(cm)  & - & 17 & 17 & 17&  17 \\
 2 & $\psi$ &  
 & $1.30_{-0.02}^{+0.02}$ 
 & $ 1.39_{-0.02}^{+0.02}$ 
 &  $1.53_{-0.04}^{+0.04}$
 & $1.61_{-0.06}^{+0.06}$ 
 & $ 1.37_{-0.03}^{+0.03}$ \\
 \\
 3 & $\kappa$ &  
 & $0.157_{-0.004}^{+0.004}$
 & $0.144_{-0.004}^{+0.004}$ 
 & $0.128_{-0.003}^{+0.003}$ 
 & $0.120_{-0.003}^{+0.003}$ 
 & $ 0.146_{-0.004}^{+0.004}$ \\
 \\
 4 & N & $10^{-8}$  
 & $7.49_{-1.0}^{+1.0}$ 
 & $24.70_{-8.0}^{+8.0}$ 
 & %$17.9_{-0.2}^{+0.2}$ & 
% $6.878e^{-7}$ 
$209.0_{-50.0}^{+50.0}$
 %& $37.4_{-0.6}^{+0.6}$ 
& $897.3_{-200.0}^{+200.0}$ 
 & $20.06_{-5.0}^{+5.0}$ \\
 \\
 5 & $log \ P_{jet}$ &  & $44.1847_{-0.007}^{+0.007}$ &
 $44.3757 _{-0.01}^{+0.01}$  & $44.6721_{-0.1}^{+0.1}$  & $44.9290_{-0.1}^{+0.1}$ &$44.3367_{-0.008}^{+0.008}$\\\\
% 5 & $P_{jet}$ & $10^{44}$ erg/$sec$ &  $1.85_{-0.03}^{+0.03}$ &  $2.37_{-0.05}^{+0.05}$  & $ 3.18_{-0.73}^{+0.73}$  & $ 7.23_{-1.67}^{+1.67}$ & $2.38_{-0.04}^{+0.04}$\\\\
 6 & $\chi^{2}(dof)$ &  
 & 365(242) 
 & 80.6(85)
 & 130.5(107) 
 & 101.5(106)  
 & 137.0(95) \\
 \\
% 7 & $factor_{magic}$ &  &  $1.2_{-0.1}^{+0.1}$  & $1.3_{-0.4}^{+0.4}$ & $1.7_{-0.5}^{+0.5}$ & $2.0_{-0.5}^{+0.5}$ & $1.3_{-0.1}^{+0.1}$

 7 & $factor_{sxt}$ &  
 & $0.93_{-0.04}^{+0.04}$ 
 & $0.94_{-0.06}^{+0.06}$ 
 & $0.74_{-0.04}^{+0.04}$ 
 & $0.93_{-0.04}^{+0.04}$
 & $0.68_{-0.04}^{+0.04}$ \\
 \\

\hline \hline
\multicolumn{8}{c}{EDA model ($constant*redden*TBabs*eblcor*sscicon*eda$)} \\
\hline \hline

1 & $B$ & $10^{-3}$ G 
 & $1.26_{-0.1}^{+0.1}$ 
 & $1.28_{-0.1}^{+0.1}$ 
 & $2.75_{-0.2}^{+0.2}$ 
 & $3.15_{-0.2}^{+0.2}$ 
 & $1.24_{-0.1}^{+0.1}$ 
 \\
 %2 & $R$ & log(cm)  & 17 & 17.00 & 17& 17 & 17 \\
\\
 2 & $\kappa$ &  
 & $0.144_{-0.004}^{+0.004}$ 
 & $0.132_{-0.004}^{+0.004}$ 
 & $0.120_{-0.005}^{+0.005}$ 
 & $0.111_{-0.003}^{+0.003}$
 & $0.135_{-0.004}^{+0.004}$ \\
 \\
 3 & $\psi$ &  & 
 $1.46_{-0.03}^{+0.03}$ 
 & $1.53_{-0.02}^{+0.02}$
 & $1.65_{-0.04}^{+0.04}$ 
 & $1.74_{-0.06}^{+0.06}$
 & $1.50_{-0.03}^{+0.03}$\\
 \\
 4 & N & $10^{-7}$  
 & $4.47_{-0.9}^{+0.9}$
 & $16.90_{-7.0}^{+7.0}$ 
 %& $54.4_{-0.2}^{+0.2}$
 & $130.0_{-40.00}^{+40.00}$
 & $814.0_{-200.00}^{+200.00}$
 & $11.30_{-5.0}^{+5.0}$ \\
 \\

\hline
\end{tabular} 
}
\\
{\bf Notes}: For broadband analysis, the size of the emission region we have considered is $R = 10^{17}$ cm and bulk Lorentz factor $\Gamma = 20$. \\
Jet Power $P_{jet}$ is in logarithmic scale with the units of erg $s^{-1}$.
\end{table*}

\addtocounter{table}{-1}
\begin{table*}
%\ContinuedFloat
\centering
%\footnotesize
\caption{(Continued...) The best-fitted spectral parameter values of the log-parabola, broken power law, PL with $\gamma_{max}$, EDD, and EDA models. The broadband SED fitting for the various models is constructed using observations from UV to gamma rays at each epoch.}
\label{tab:sed_fitpar_cont2}
\vspace{0.3cm} 
\resizebox{\textwidth}{!}{ %

\begin{tabular}{lccccccr}

\hline \hline
S.N.  & Model  & Unit & epoch-1 & epoch-2 & epoch-3 & epoch-4 & epoch-5 \\ %& epoch-6\\
 & Parameters & & 21--23 Sep 17 & 9--10 Dec 17 & 21--22 Dec 17 & 8--9 Jan 18 & 8--12 Aug 21\\
\hline \hline

 5 & $log \ P_{jet}$ &  & 
 $44.1053_{-0.007}^{+0.007}$ &$ 
 44.0360_{-0.01}^{+0.01}$ &
 $44.5188 _{-0.01}^{+0.01}$  
 & $44.8804 _{-0.01}^{+0.01}$&
 $44.23_{-0.008}^{+0.008}$\\
 \\
% 5 & $P_{jet}$ & $10^{44}$ erg/$sec$ &  $1.57_{-0.02}^{+0.02}$ &$ 1.44_{-0.03}^{+0.03}$& $2.63_{-0.06}^{+0.06}$  & $4.24_{-0.09}^{+0.09}$ & $1.89_{-0.03}^{+0.03}$\\
 \\

6 & $\chi^{2}(dof)$ &  
 & 367(242) 
 & 82.3(88) 
 & 131(107)
 & 101.7(106)  
 & 137.8(95) \\
 \\
% 7 & $factor_{magic}$ &   & $1.24_{-0.1}^{+0.1}$ & $1.26_{-0.1}^{+0.1}$  & $1.34_{-0.6}^{+0.6}$ & $1.56_{-0.3}^{+0.3} $ & $1.34_{-0.2}^{+0.2}$ \\
 \\
% 9  & $factor_{sxt}$ &  & $0.91_{0.04}^{0.04}$ & $0.93_{0.03}^{0.03}$ & $0.73_{0.05}^{0.05}$ & 1.41  & $0.67_{0.03}^{0.03}$  \\
 7 & $factor_{sxt}$ &  
 & $0.92_{-0.04}^{+0.04}$ 
 & $0.94_{-0.06}^{+0.06}$ 
 & $0.74_{-0.04}^{+0.04}$ 
 & $0.93_{-0.04}^{+0.04}$
 & $0.67_{-0.04}^{+0.04}$ \\
 \\

\hline
\end{tabular} 

}
\\
\textbf{Notes}:  For broadband analysis, the size of the emission region we have considered is R= $10^{17}$ cm and bulk Lorentz factor, $\Gamma$ = 20. Jet Power $P_{jet}$ is in logarithmic scale with the units of erg $s^{-1}$. 
$factor_{sxt}$ is the relative cross-normalization constant between X-ray instrument SXT and LAXPC. This factor was kept frozen at 1 for LAXPC, whereas it was kept free for the SXT instrument.
\end{table*}

%-----------------------------------------------------------------------------------------------------------------------------------

For this study, we utilize empirical models, such as log parabola (LP) and broken power law (BPL), and we input the particle distributions to the \emph{$sscicon \otimes n(\xi)$} model. Additionally, we incorporate physical models, including power-law with maximum energy due to radiative cooling (PL with $\gamma_{max}$), EDD, and EDA in order to fit the observed spectrum with the SSC emission. The goal of this analysis is to comprehensively explore and understand the observed spectrum through the application of both empirical and physical models.

%For this study, the input particle distributions to the \emph{$sscicon \otimes n(\xi)$}  model are obtained using empirical models, like log parabola (LP) and broken power law (BPL). Additionally, we have used physical models such as power-law with maximum energy due to radiative cooling, energy-dependent diffusion (EDD), and energy-dependent acceleration (EDA) to fit the observed spectrum with the SSC emission.
\setlength{\leftmargini}{10pt}
\begin{itemize}
    \item {\bf Log parabola (LP)}: The underlying particle distribution in this scenario is described as  
\begin{equation}
 \label{eq:logpar_conv}
     n(\xi)  = K ({{\xi}/{\xi_{r}}})^{-{\alpha}-{\beta}{\log({{\xi}/{\xi_{r}}})}}\\
\end{equation}

here, particle spectral index is denoted as ${\alpha}$ at the reference energy $\xi^2=\xi_r^2$,  the spectral curvature is represented by ${\beta}$, and K is the normalization of the particle density. The reference energy $\xi_r^2$ kept constant at one keV throughout the spectral fit, whereas the parameters $\alpha$, $\beta$, and norm K remained free.

\item {\bf Broken power-law (BPL)}: This scenario of particle distribution is given by
%\textbf{
\begin{equation}
    \label{eq:bknpo}
    n(\xi)= \begin{cases}
        K (\xi/1 \sqrt{keV})^{-p} \ \text{for} \ \xi < \xi_{break} \\
        K \xi^{q-p}_\text{break}(\xi/1\sqrt{keV})^{-q}  
    \ \text{for} \ \xi > \xi_{break}
    \end{cases}
\end{equation}
%}
Where K represents the normalization, $p$ and $q$ represent the low and high energy photon indices, respectively, while the Lorentz factor associated with the break energy is $\xi_{break}$ and the transformation is represented by $\xi$=$\gamma \sqrt{\mathbb{C}}$.
The distributions is non-zero for $\xi_{min} < \xi < \xi_{max}$, which correspond to $\gamma_{min}$ and $\gamma_{max}$ such that $\xi_{min}$=$\gamma_{min} \sqrt{\mathbb{C}}$ and $\xi_{max}$=$\gamma_{max} \sqrt{\mathbb{C}}$.
%the distributions is non-zero for \xi  between \xi_min and \xi_max, which correspond to gamma_min and gamma_max

%.  The minimum and maximum Lorentz factors of the electron are $\xi_{min}$ and $\xi_{max}$, respectively, while the Lorentz factor associated with the break energy is $\xi_{break}$.

\item {\bf Power-law particle distribution with maximum electron energy (PL with $\gamma_{max}$)}: In this case, we have considered that the particles accelerated by a shock, and then subsequently, these accelerated particles lose energy through radiative processes.

In such a case, the steady-state particle density is given by
\begin{equation}\label{kinetic}
\frac{\partial}{\partial \gamma}\left[\left(\frac{\gamma} {\tau_{acc}}-\beta_s\gamma^2\right)n_a\right]+\frac{n_a}{\tau_{esc}}=Q\delta(\gamma-\gamma_0)
\end{equation}
where $\tau_{acc}$ and $\tau_{esc}$ represent the acceleration and escape time scales, respectively, and the radiative loss term includes $\beta_s=\frac{4}{3}\frac{\sigma_TB^2}{8\pi m_ec}$,  B is the magnetic field, $\sigma_T$ is the Thomson cross-section and $m_e$ the electron mass. We consider that a mono-energetic injection of electrons Q at energy $\gamma_0$. The solution of the steady-state equation is described as
%the blob populated by power-law particle distribution with maximum electron energy ($\xi_{max}$) electron distribution described as
\begin{equation}
  \label{eq:gmax_part}
  n(\xi) = K \xi^{-p} \left(1-\frac{\xi}{\xi_{max}}\right)^{(p-2)}
\end{equation}
Where, $K=Q_0\tau_a\gamma_0^{p-1}\mathbb{C}^{p/2}$, $p= \tau_{acc}/\tau_{esc} +1$  is the particle spectral index, $\xi_{max}=\gamma_{max}\sqrt{\mathbb{C}}$ with $\gamma_{max}  = 1/(\beta_s \tau_{acc})$ is the maximum Lorentz factor that an electron can attain before losing energy.
The free parameters are,  $\xi_{max}$, $p$,  and the normalization  $\mathbb{N}$ defined as
\begin{equation} \label{eq:norm_ximax}
\mathbb{N}  = \frac{\delta^3(1+z)}{d_L^2}V\,\mathbb{A} Q_0\tau_{acc}{\gamma_0^{p-1}} \mathbb{C}^{p/2}
\end{equation}
 
For $\gamma_{max}$ model, the distribution is only for $\xi$$ > $$\xi_{min}$ such that $\xi_{min}$=$\gamma_{min} \sqrt{\mathbb{C}}$.

\item {\bf{Energy dependent diffusion model (EDD)}}: In this case, we assume that the diffusion takes place in a region consisting of the tangled magnetic field, which may cause the diffusion coefficient dependent on the gyration radius. Consequently, escape time scale energy dependent or the diffusion coefficient energy is dependent as $\tau_{esc}({\gamma}$) is given by $\tau_{esc} = \tau_{esc,R} \left(\frac{\gamma}{\gamma_R}\right)^{-\kappa} $. %In this case the underlying particle distribution is assumed  by considering energy dependence of the escape time-scale
By ignoring the synchrotron energy loss  and  considering  this escape time-scale dependence, the solution to \autoref{kinetic} will be  \citep[for detailed derivation see][]{2021HOTA, 2022Khatoon}
\begin{equation}
  n(\xi)=Q_o \tau_{acc}\sqrt{\mathbb{C}} \xi^{-1}\rm{exp}\left[-\frac{\eta_R}{\kappa}\,\left(\left(\frac{\xi}{\xi_R}\right)^{\kappa}-\left(\frac{\xi_0}{\xi_R}\right)^{\kappa}\right)\right]
  \label{n_EDDt}
\end{equation}
where $\xi_R = \sqrt{\mathbb{C}}\gamma_R$, $\xi_0 = \sqrt{\mathbb{C}}\gamma_0$ and $\eta_R \equiv \tau_{acc}/\tau_{esc,R}$. 

After removing the degenerate parameters, we can use the updated Equation as follows

\begin{equation}
\label{eq:eddcr1}
	n(\xi)=K^\prime \xi^{-1}\rm{exp}\left[-\frac{\psi}{\kappa}\,\xi^{\kappa}\right]
\end{equation}

where $\psi=\eta_R\left(\mathbb{C}\gamma_R^2\right)^{-\kappa/2} = \eta_R \xi_R^{-\kappa} $, and the normalization parameter ($K^\prime$) is given as % in \autoref{eq:edd_kdash} which is further modified into $\mathbb{N}$ (\autoref{eq:norm_edd})

\begin{equation}
\label{eq:edd_kdash}
K^\prime = Q_0 \tau_{acc} \sqrt{\mathbb{C}} \rm{exp}\left[\frac{\eta_R}{\kappa}\,\left(\frac{\gamma_0}{\gamma_R}\right)^{\kappa}\right]
\end{equation}

and further modified into $\mathbb{N}$ as

\begin{equation}\label{eq:norm_edd}
   \mathbb{N}  = \frac{\delta^3(1+z)}{d_L^2} V \mathbb{A}K^\prime
\end{equation}

We have considered $\psi$, $\kappa$, and $\mathbb{N}$ as the free parameters for the above model.

\item {\bf{Energy dependent acceleration model (EDA)}}: We next consider a case where the energy dependence of acceleration time scale as $\tau_{acc} = \tau_{acc,R}\left(\frac{\gamma}{\gamma_R}\right)^\kappa$. So considering the above dependency $\tau_{acc}$, the solution to \autoref{kinetic} will be \citep[defined in][]{2021HOTA,2022Khatoon} 
\begin{equation}\label{eq:edatxi}
%	n(\xi)=K \frac{\xi^{\kappa-1}}{\mathbb{C}^{(\kappa-1)/2}} \rm{exp}\left[-\frac{\psi}{\kappa}\xi^{\kappa}\right]
n(\xi)=Q_0\tau_{acc,R}\sqrt{\mathbb{C}}\xi_R^{-\kappa}\xi^{\kappa-1}\exp \left[-\frac{\eta_R}{\kappa}\left(\left(\frac{\xi}{\xi_R}\right)^\kappa-\left(\frac{\xi_0}{\xi_R}\right)^\kappa\right)\right]
\end{equation}
where $\xi_0 = \sqrt{\mathbb{C}}\gamma_0$, $\xi_R = \sqrt{\mathbb{C}}\gamma_R$ and $\eta_R \equiv \tau_{acc,R}/\tau_{esc}$.

We can recast the distribution as 
\begin{equation}
\label{eq:edacr}
	n(\xi)=K^\prime \xi^{\kappa-1}\rm{exp}\left[-\frac{\psi}{\kappa}\,\xi^{\kappa}\right]%\quad\rm{for}\quad \gamma\gg\frac{1}{Bt_{esc}}
\end{equation}
where $\psi=\eta_R\left(\mathbb{C}\gamma_R^2\right)^{-\kappa/2} = \eta_R \xi_R^{-\kappa} $, and the normalization ($K^\prime$) is defined in \autoref{eq:eda_kdash} and further modified as $\mathbb{N}$ (\autoref{eq:norm_eda}).
\begin{equation}
\label{eq:eda_kdash}
K^\prime = Q_0 \tau_{acc,R}\sqrt{\mathbb{C}}\xi_R^{-\kappa} \rm{exp}\left[\frac{\eta_R}{\kappa}\,\left(\frac{\xi_0}{\xi_R}\right)^{\kappa}\right]
\end{equation}

\begin{equation}\label{eq:norm_eda}
\mathbb{N}  = \frac{\delta^3(1+z)}{d_L^2} V \,\mathbb{A}K^\prime
\end{equation}

We have considered $\psi$, $\kappa$, and $\mathbb{N}$ as the free parameters for the above model.

\end{itemize}

To consider Galactic absorption while fitting the broadband SED of the source, we used the TBabs model \citep{2000ApJ...542..914W} available in the XSPEC with the equivalent hydrogen column density ($N_{H}$) set at $7.9 \times 10^{20}$cm$^{-2}$  as determined by the online tool{\footnote{https://heasarc.gsfc.nasa.gov/cgi-bin/Tools/w3nh/w3nh.pl}} created by the LAB survey group \citep{2005A&A...440..775K}.

%+++++++++++++++++++New Ranjeev+++++++++++++++

For the SED spectral fitting, we fixed the size R, the bulk Lorentz factor $\Gamma$, and the opening angle $\theta$, at $10^{17}$ cm, 20, and 0$^\circ$,
respectively, while keeping other parameters free. However, we have discussed later the impact of variations of these parameters.
For the broken power-law model, we further fix the minimum and maximum Lorentz factors, $\gamma_{min}$ and $\gamma_{max}$ to 10 and $10^8$, respectively.
The best-fit parameters with errors are listed in \autoref{tab:sed_fitpar} along with $\chi^2$ for each of the epochs and different particle distributions.
\autoref{fig:sed_epoch2} shows the model along with the data for a representative epoch. 
The solid lines in \autoref{fig:sed_epoch2} represent the best-fit model without Galactic absorption. The dotted lines represent the best-fit model spectra for the SXT data with Galactic absorption. Note that to take into account the calibration uncertainties between  LAXPC and SXT instruments a constant factor has been multiplied to the SXT model, which causes the dotted lines to be shifted compared to the solid ones. The constant factor was fixed for LAXPC, whereas it was free to vary for the SXT ($factor_{sxt}$) instrument. The constant factor for SXT ($factor_{sxt}$) for each of the epochs is listed in \autoref{tab:sed_fitpar}.

\subsection{Jet Power}
\label{sec:jet}

We have calculated the total jet power ($P_{jet}$), or the total power carried by electrons, Poynting flux, radiation, and protons \citep{2008Celotti} using \autoref{eq:jet_power} as given below, 
 \begin{equation}
 \label{eq:jet_power}
     P_{jet}= 2 \ \pi^2 \ R^2 \ \Gamma^2 \ \beta_c \  u^\prime_k
 \end{equation}
Here, factor 2 refers to two-sided jets, $\Gamma$ represents the bulk Lorentz factor, R is the size of the emission region, and $u^\prime_k$ are the energy densities in the co-moving jet’s frame of the magnetic field (k = mag), relativistic electrons (k = ele), and cold protons (k = kin). %To calculate the jet power, we freeze the value of R and $\Gamma$ at $10^{17}$ cm and 20, respectively. We have estimated the jet power for all five different particle distribution models, and the values are given in \autoref{tab:sed_fitpar}. 

\begin{figure}
  \centering
        \includegraphics[width=\columnwidth]{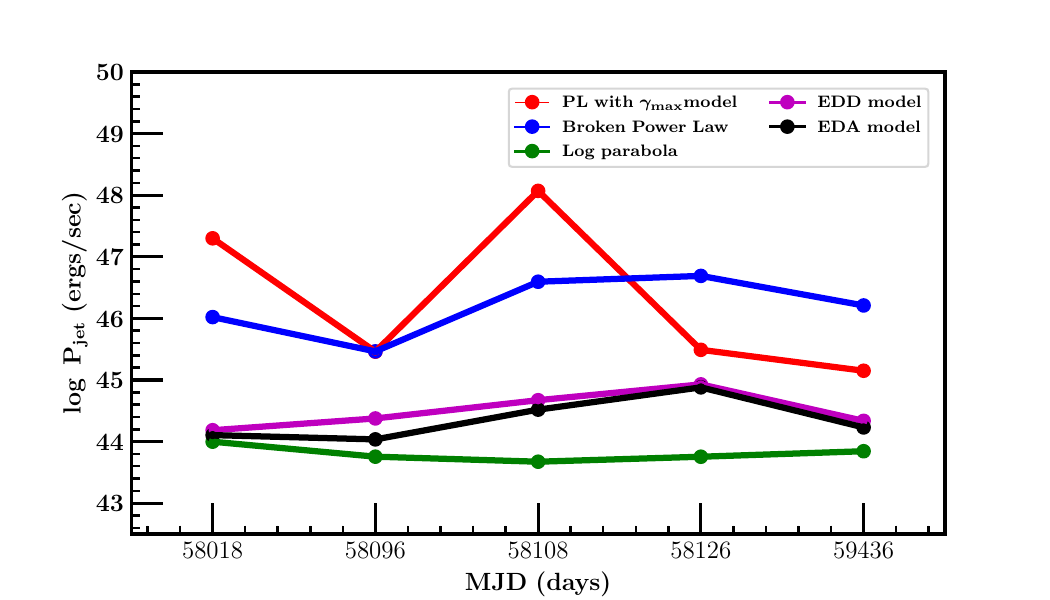}
   \caption{Jet power estimated for all five models as indicated in the legend at five different epochs.}
   \label{fig:jet_power}
\end{figure} 

We have calculated the $P_{jet}$ values for all the five particle distribution models used for the broadband SED fitting described in Section \ref{sec:sedmodeling}. Since we have performed SED fitting by freezing R and $\Gamma$ values at $10^{17}$ cm and 20, respectively, $P_{jet}$ values are calculated for the fixed R and $\Gamma$ for all the particle distribution models and at all the five epoch observations, given in \autoref{tab:sed_fitpar}.
In \autoref{fig:jet_power}, we present the variation of jet power at different epochs for all the models. The maximum jet power values are obtained for the PL with $\gamma_{max}$ model and the broken power law model i.e., approximately two orders of magnitude higher compared to the log parabola, EDD, and EDA models. Notably, the values of jet power remain nearly constant for the log parabola, EDD, and EDA models, while for the broken power law and PL with $\gamma_{max}$ models, the jet power values exhibit variability across all five epochs.

\begin{figure*}	
\centering
\includegraphics[width=\columnwidth]{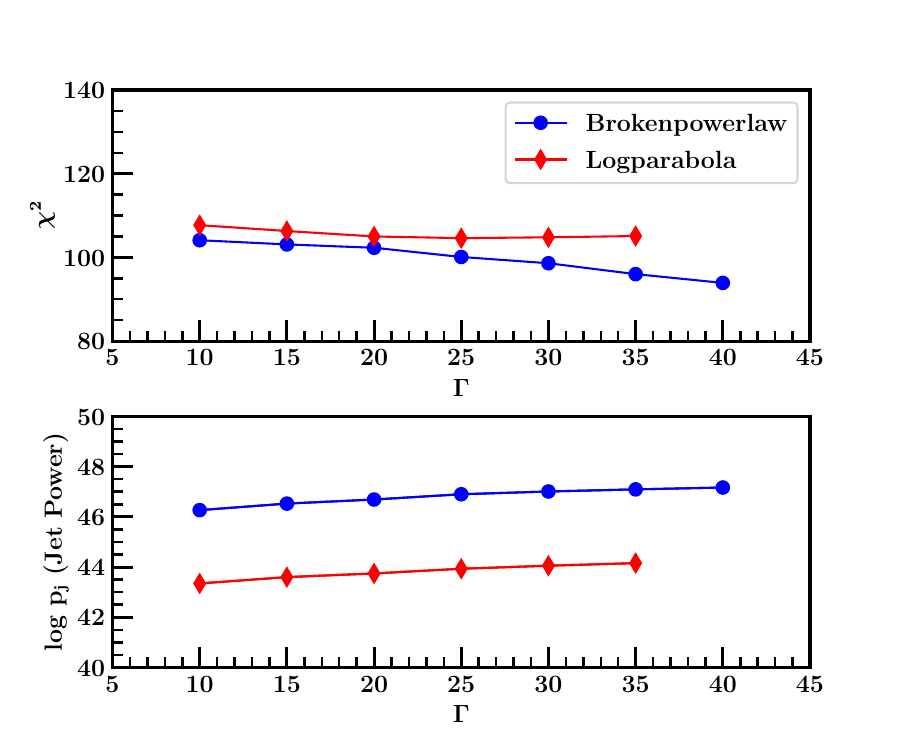}
\includegraphics[width=\columnwidth]{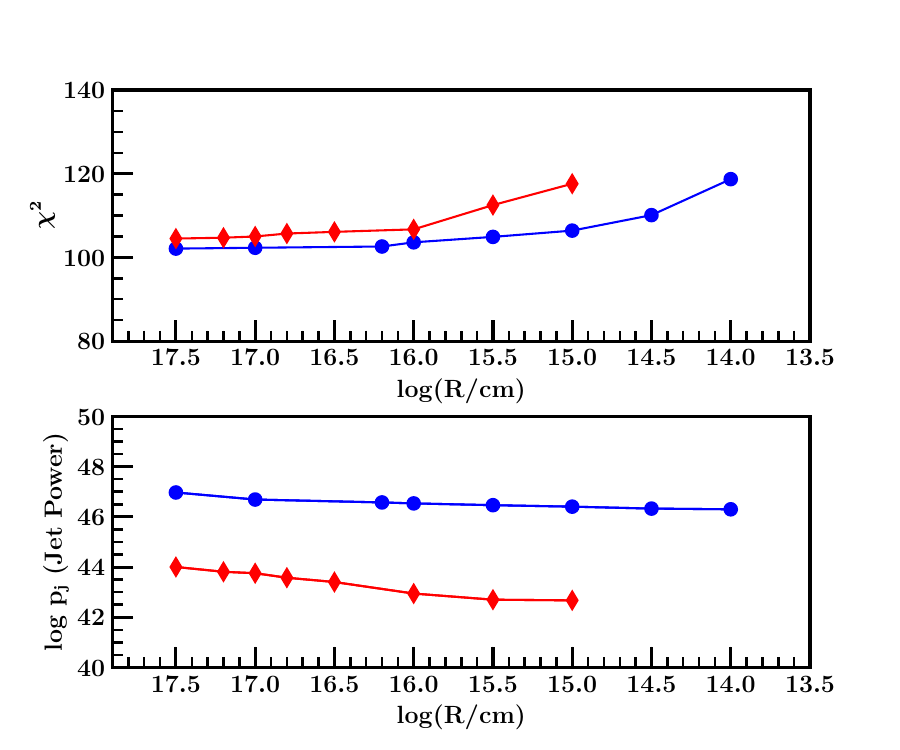}
\includegraphics[width=\columnwidth]{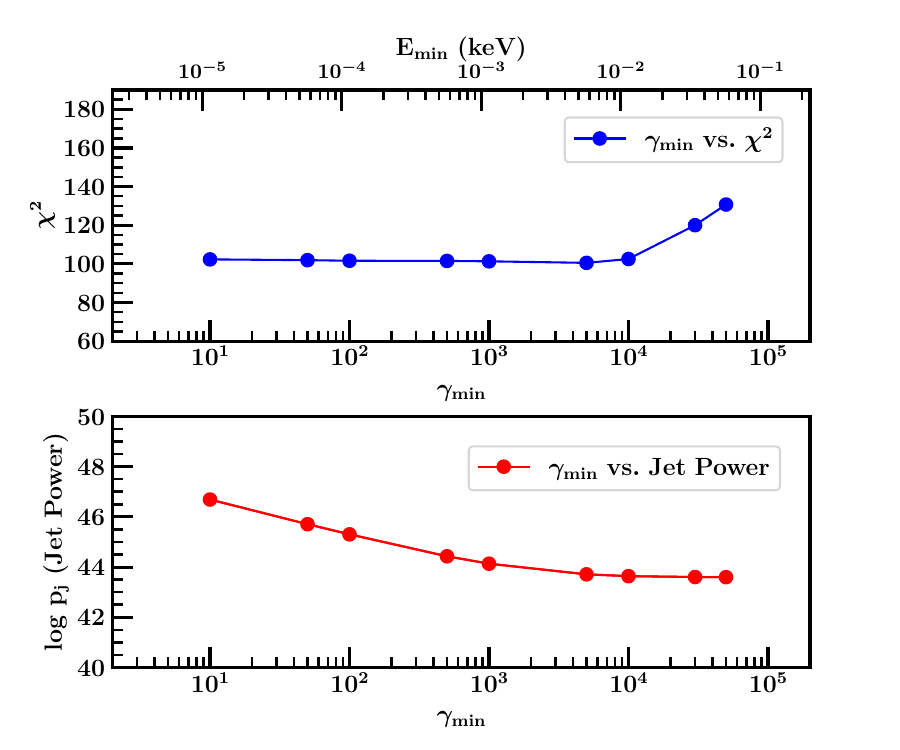}
\caption{Top left panel represents the variation of $\Gamma$ with $\chi^2$ and $P_{jet}$ whereas the top right panel represents the variation of size of the region (R) with $\chi^2$ and $P_{jet}$ for both the Broken power law (blue) and Log parabola (red) models for data epoch-4. The bottom panel shows the variation of $\gamma_{min}$ with $\chi^2$ and $P_{jet}$  only for the Broken power law model.}
\label{fig:gamma_vs_jetpower}
\end{figure*}

Furthermore, to examine the variation in the jet power of the blazar source 1ES 0229+200 at different values of $\Gamma$ and R, we calculated the jet power by systematically varying 
%$\Gamma$ (upper panel) and R (middle panel)
$\Gamma$ and R for both the log parabola and broken power law models.
The top left panel of  \autoref{fig:gamma_vs_jetpower} illustrates the variation of $\Gamma$  with $\chi^2$ (top) and jet power (bottom). The jet power shows a slight increase (by a factor of one or two) as the $\Gamma$ values vary from 10 to 35, becoming approximately constant at $\Gamma \geq 35$. In the top right panel of  \autoref{fig:gamma_vs_jetpower}, the variation of R with its respective $\chi^2$ (top) and jet power (bottom) values is depicted. The plot indicates a slight decrease (by a factor of 2(1)) in jet power for the log parabola model (broken power law), and the $\chi^2$ remains stable with decreasing R values. However, $\chi^2$ values %become unstable 
increases below R = $10^{15}$ ($10^{16}$) for the broken power-law (log parabola model). Therefore, the general $P_{jet}$ estimates with $\Gamma$ = 20 and size R = $10^{17}$ cm are insensitive to their presumptions. 

Additionally, the bottom panel of \autoref{fig:gamma_vs_jetpower}  displays the variation in jet power (bottom) and $\chi^2$ (top) of the source 1ES 0229+200 with respect to $\gamma_{min}$ for the broken power-law particle distribution. The jet power decreases with increasing $\gamma_{min}$ for large values, while the $\chi^2$ remains constant up to $10^4$ and increases beyond that threshold. We find that the $P_{jet}$ for the broken power-law model is $\sim$ $10^{44}$ ergs/sec for a value as large as $\gamma_{min}$ = $10^4$.

%Further we have calculate the Jet power at different radii.

%------------------------------- Corelation plots ------------------------------------

\subsection{Correlation study of spectral parameter }
\label{sec:corel}

The study of correlation among various spectral parameters is important to understand the dependence between the fit parameters and the observed properties. We obtained spectral parameters for log-parabola, Broken power law, PL with $\gamma_{max}$, EDD, and EDA models. Then, we determined Spearman's rank correlation coefficient ($r_s$) and null hypothesis probability ($P_{rs}$) for all the derived parameters in each model.

\begin{figure*}
  \centering
 	\includegraphics[width=0.3\textwidth]{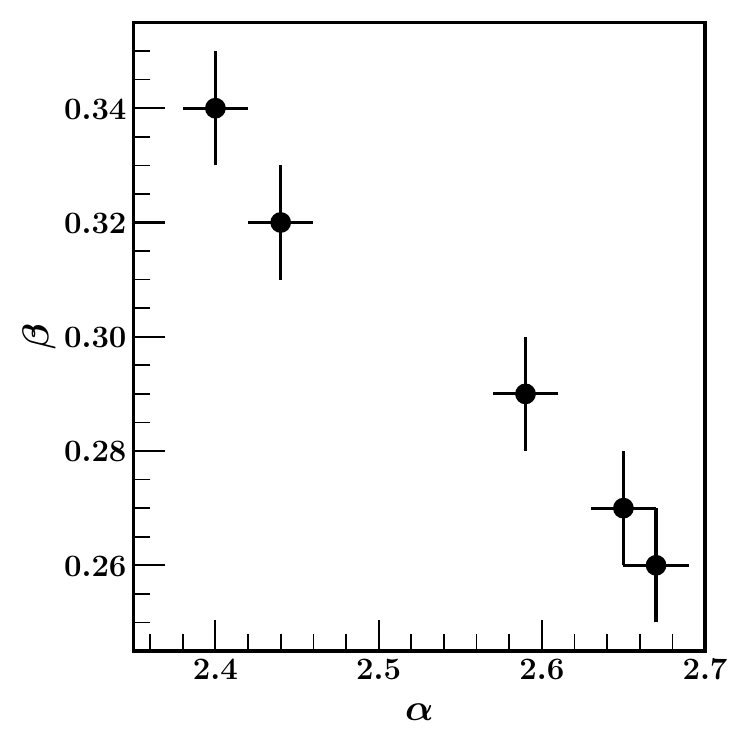}
    \includegraphics[width=0.3\textwidth]{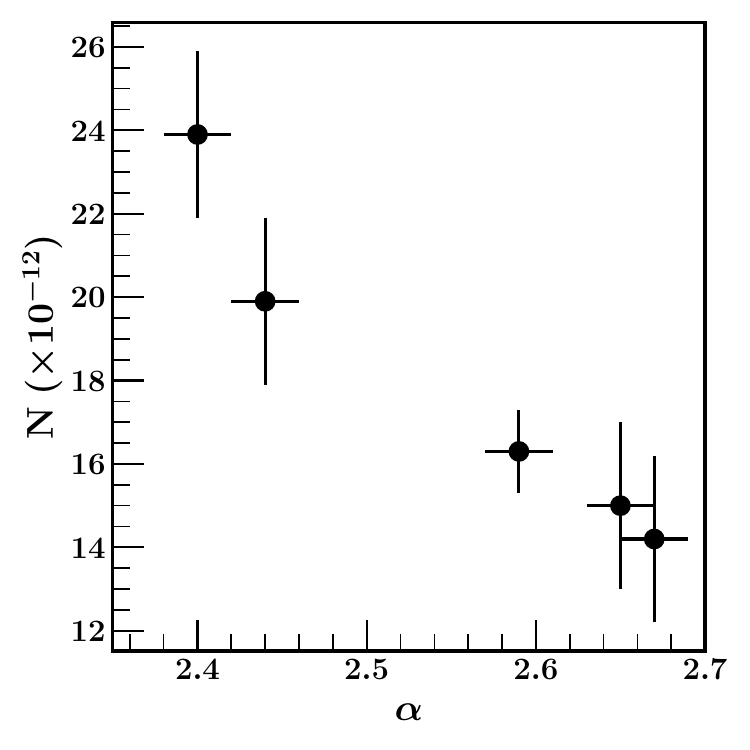}
  	\includegraphics[width=0.3\textwidth]{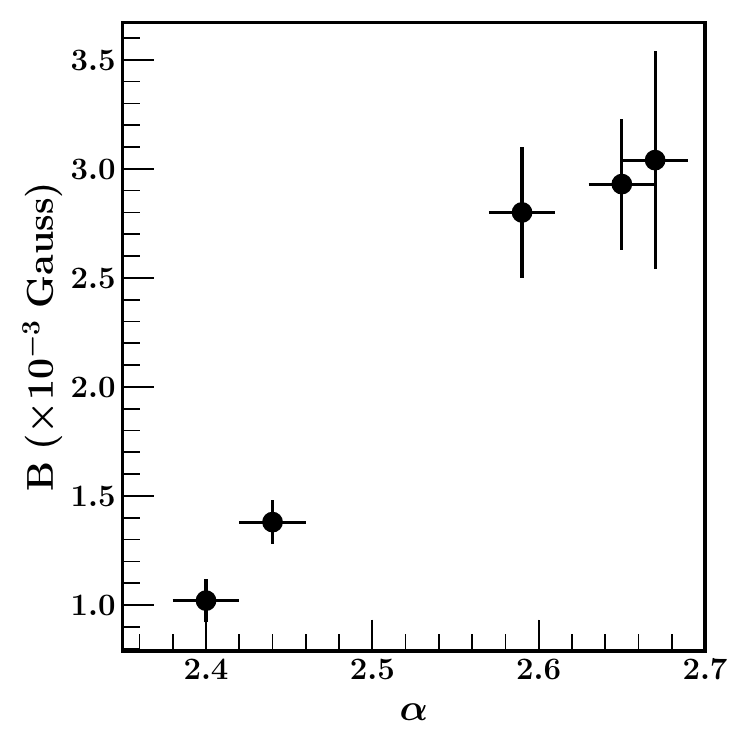}
    \includegraphics[width=0.3\textwidth]{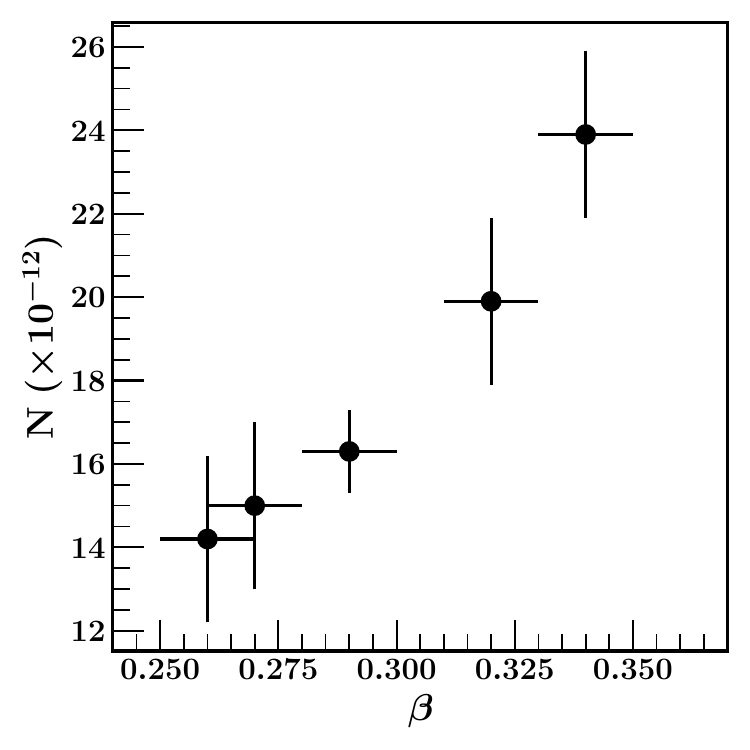}
    \includegraphics[width=0.3\textwidth]{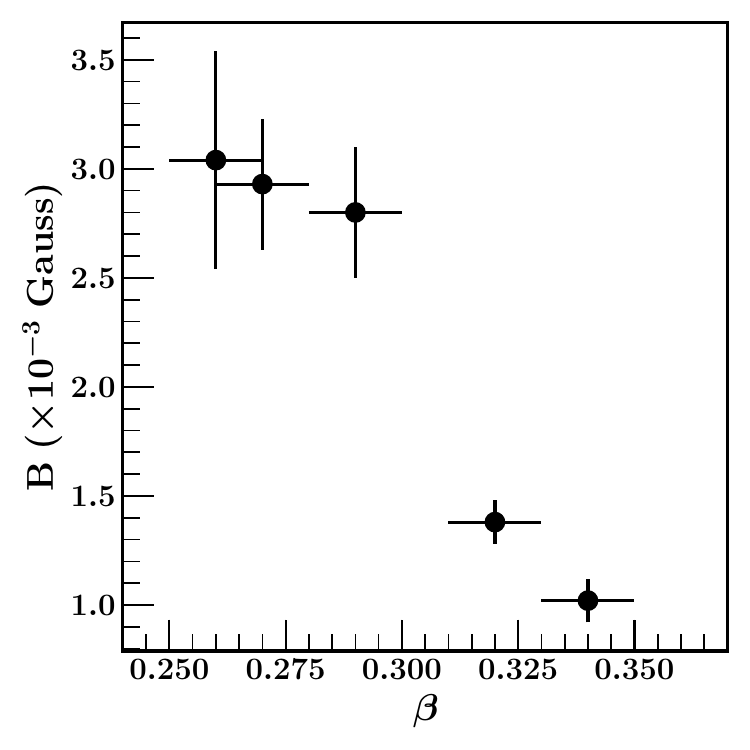}
      \includegraphics[width=0.3\textwidth]{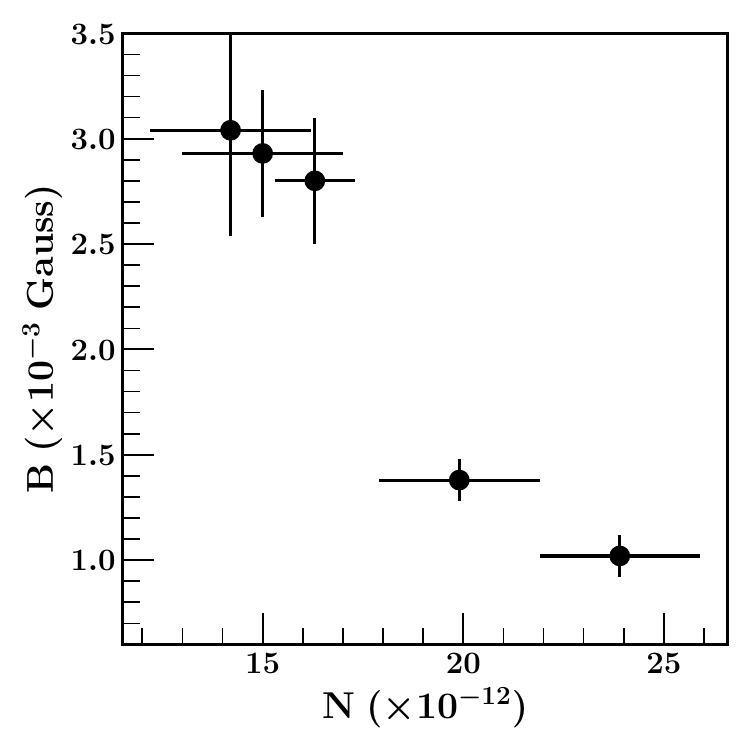}

   \caption{Scatter plots between the derived parameters obtained for log parabola model at five epochs. Top left panel: $\alpha$ vs. $\beta$, top middle panel: normalisation vs. $\alpha$, top right panel: $\alpha$ vs. magnetic field value, bottom left panel: $\beta$ vs. normalisation, bottom middle panel: $\beta$ vs. magnetic field, and bottom right panel: normalisation vs. magnetic field.
   }
   \label{fig:logpara_cor}
\end{figure*}

The scatter plots between spectral parameters of the log-parabola are shown in \autoref{fig:logpara_cor}. % and the top rows of \autoref{} provide the correlation results. 
%There is a significant anti-correlation observed 
A significant anti-correlation is observed between fit parameters,  $\alpha$ vs. $\beta$ and normalization each with $r_s= -0.99$ ($P_{rs}= 1.4\times 10^{-24})$. 
%The anti-correlation between $\alpha$ and normalization reflects the ‘Harder when brighter’ features in the spectrum frequently seen in blazars \citep{2021Giommi}. 
A positive correlation is found between $\beta$ vs. normalization and $\alpha$ vs. B, each with $r_s$= 0.99 ($P_{rs}= 1.4\times 10^{-24})$. Furthermore, a significant anti-correlation is observed between fit parameter magnetic field vs. $\beta$ and normalisation, each having $r_s = -$0.99 ($P_{rs}= 1.4\times 10^{-24})$.

%There is a significant anti correlation observed between fit parameter $\alpha$ and $\beta$ with $r_s= -0.99$ ($P_{rs}= 0.00$) and $\alpha$ verus magnetic field B with $r_s= -0.99$ ($P_{rs}= 0.00$). Similarly, the correlation between $\alpha$ and normalization is found $r_s= -0.99$ ($P_{rs}= 0.00$), reflecting the ‘Harder when brighter’ features in the spectrum frequently seen in blazars {\bf ref?}. On the other hand, a positive correlation is found between $\beta$ and normalization with $r_s$= 0.99 ($P_{rs}= 0.00$). Furthermore, a significant anti-correlation was observed between fit parameter magnetic field versus $\beta$ and Normalisation, each having $r_s$= -0.99 ($P_{rs}= 0.00$).

\begin{figure*}
  \centering
\includegraphics[width=0.3\textwidth]{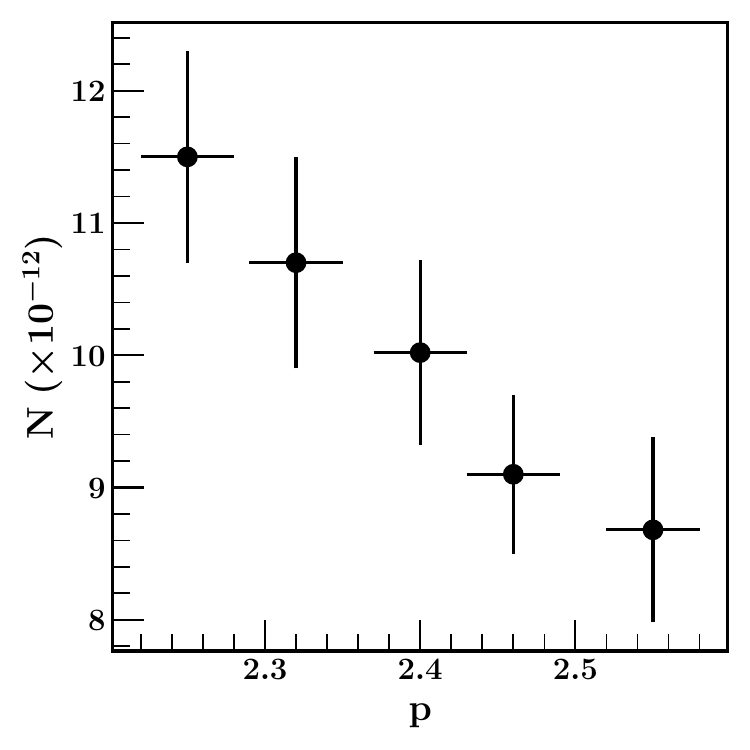}
\includegraphics[width=0.3\textwidth]{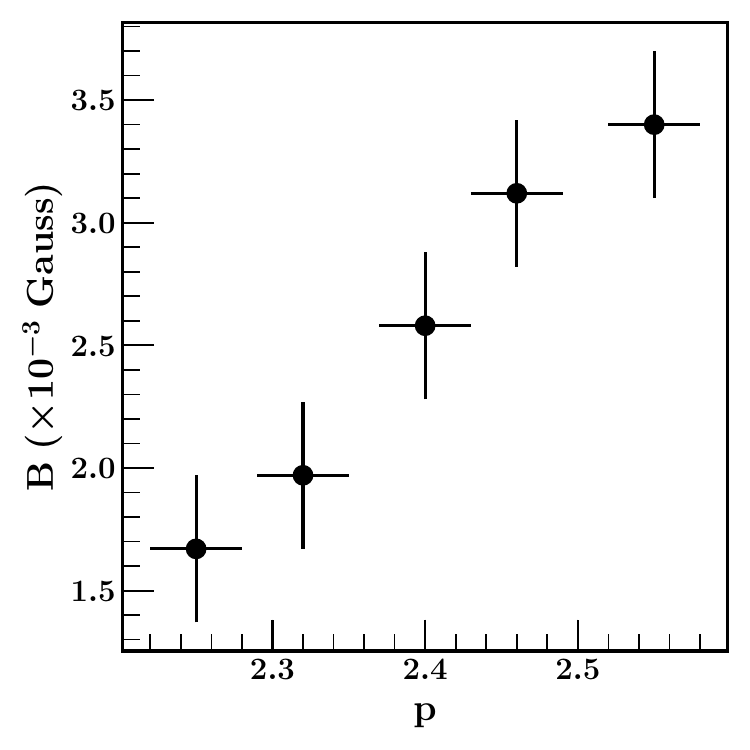}
\includegraphics[width=0.3\textwidth]{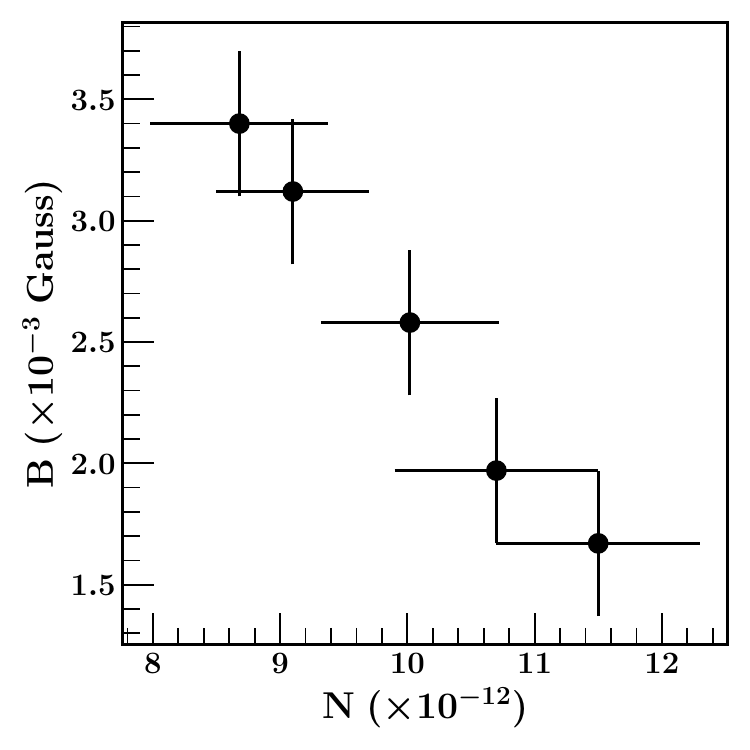}

   \caption{Scatter plots between the derived parameters with broken power law model at five epochs. Left panel: $p$ vs. normalisation,  middle panel: $p$ vs. magnetic field, and right panel: normalisation vs. magnetic field.}
   \label{fig:bknpo_cor}
\end{figure*} 

\begin{figure*}
  \centering
    \includegraphics[width=0.3\textwidth]{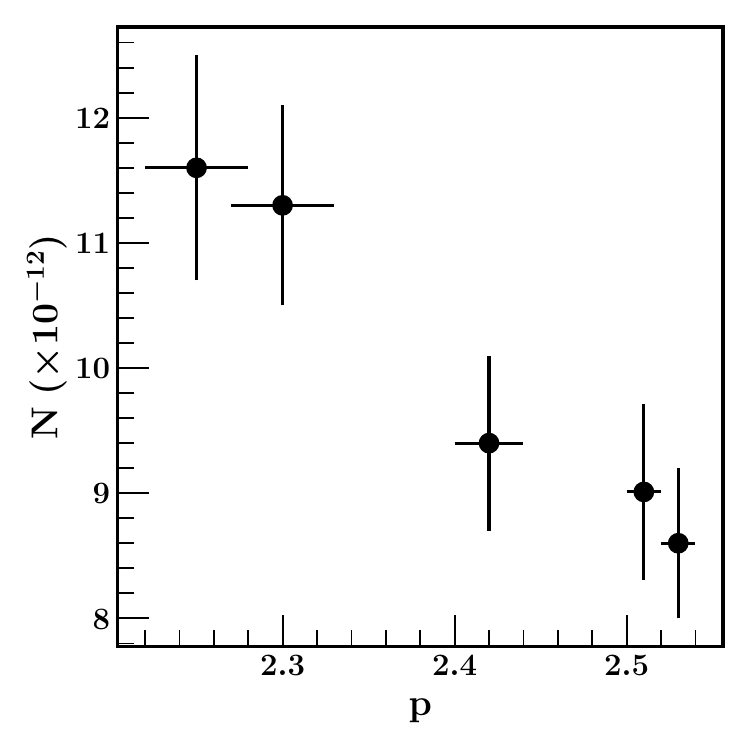}
    \includegraphics[width=0.3\textwidth]{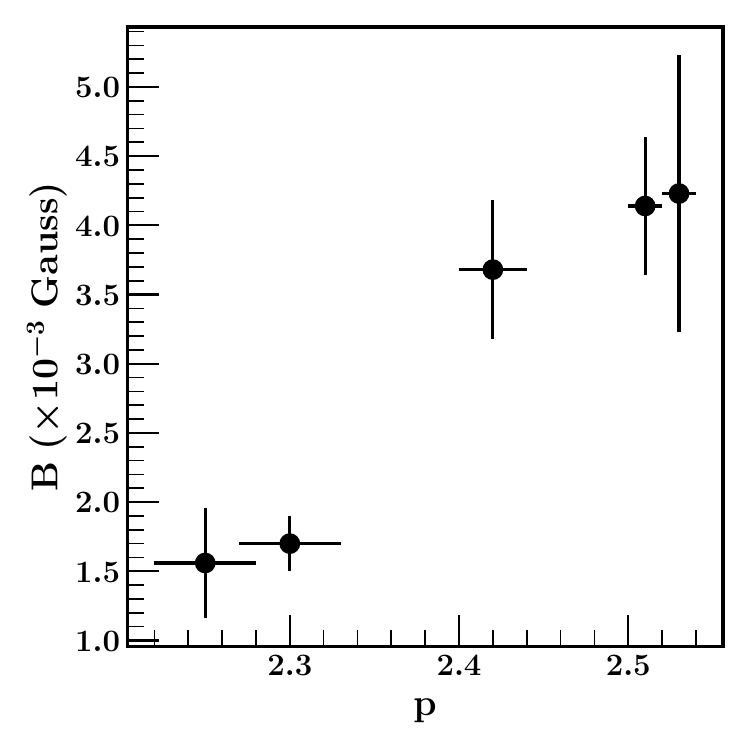}
    \includegraphics[width=0.3\textwidth]{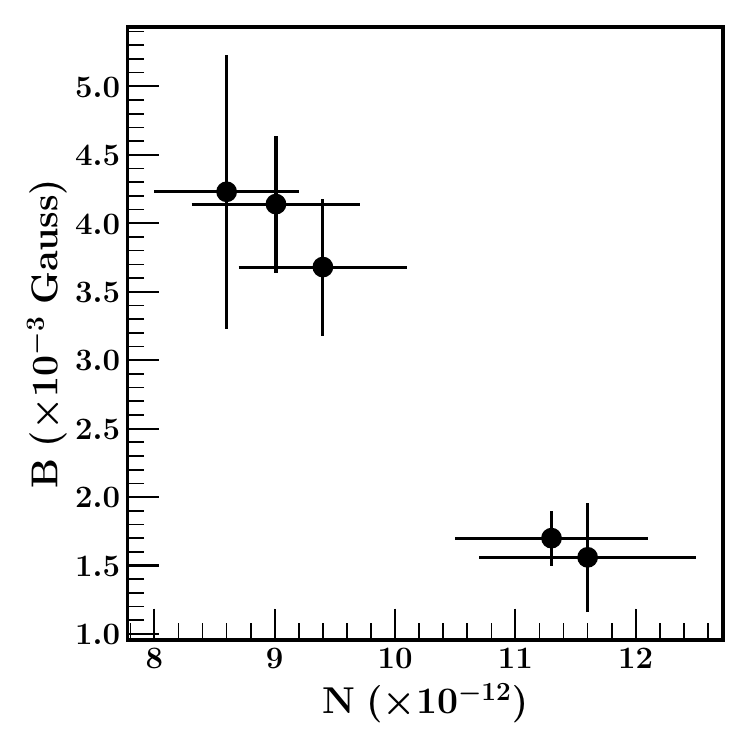}

   \caption{Scatter plots between the  derived parameters with PL with $ \gamma_{max}$ model at five epochs. Left panel: $p$ vs. normalisation,  middle panel: $p$ vs. magnetic field, and right panel: normalisation vs. magnetic field.}
   \label{fig:gammamax_cor}
\end{figure*}

The scatter plots for the broken power law model parameters are presented in \autoref{fig:bknpo_cor}.
%and the correlation values are detailed in the second rows of \autoref{}.
A strong anti-correlation is observed between index `p' vs. normalization and the magnetic field `B' vs. normalization, both with  $r_s= -0.99$ ($P_{rs}= 1.4\times 10^{-24})$. On the contrary, a strong positive correlation is found between index p and magnetic field B with $r_s= 0.99$ ($P_{rs}= 1.4\times 10^{-24})$.

%The scatter plots between the broken power law model parameters are shown in \autoref{fig:bknpo_cor}, and the correlation values between the model parameters are given in the second rows of \autoref{}. The study between index p and normalization shows a strong anti-correlation with $r_s= -0.99$ ($P_{rs}= 0.00$), and a similar result is found for normalization versus magnetic field B with $r_s= -0.99$ ($P_{rs}= 0.00$). On the contrary, a strong correlation between index p and magnetic field B is observed with $r_s= -0.99$ ($P_{rs}= 0.00$).

%The Spearman’s rank correlation results for the $\xi_{max}$ model fit parameters are presented in the third rows of \autoref{}, and 
\autoref{fig:gammamax_cor} illustrates the scatter plot for the  PL with  $\gamma_{max}$ model parameters. A pronounced anti-correlation is noted in both index `p' vs. normalization and magnetic field `B' vs. normalization, each with $r_s= -0.99$ ($P_{rs}= 1.4\times 10^{-24})$. The observed strong anti-correlation between index p and normalization contradicts the theoretical form of $\mathbb{N}$ ($log \mathbb{N}\ \propto\ p$; \autoref{eq:norm_ximax}). Additionally, a strong positive correlation is also noted between index `p' and the magnetic field; however, this conflicts with the theoretical relation, B $\propto$ $(p-1)^{-1/n}$ \citep{2022Khatoon}, which predicts a negative correlation between `p' and `B'.

%The third rows of \autoref{} include the Spearman's rank correlation results for the $\xi_{max}$ model fit parameters, and \autoref{fig:gammamax_cor} shows the scatter plot between them. A strong anti-correlation is observed in index p vs. normalization and magnetic field B vs normalization, each with $r_s= -0.99$ ($P_{rs}= 0.00$).
%and magnetic field; however, the theoretical relation, B $\propto$ $(p-1)^{-1/n}$ \citep{2022Khatoon} contradicts the observed positive correlation found between p and B.

%A strong anti-correlation is observed between index p and normalization and magnetic field B, each with $r_s= -0.99$ ($P_{rs}= 0.00$). Nevertheless, the strong anti-correlation between index p and normalization contradicts the theoretical form of $\mathbb{N}$ ($log  \mathbb{N}\ \propto\ p$; \autoref{eq:norm_ximax}). Further, a strong positive correlation is noticed between index p and magnetic field, whereas B $\propto$ $(p-1)^{-1/n}$ \citep{2022Khatoon} which is contradicting to the observed positive correlation found between p and B.

%a strong anti-correlation is observed between index p and normalization and magnetic field B, each with $r_s= -0.99$ ($P_{rs}= 0.00$).

%However, when we take into account the following factors, the model may be in accordance with the observation.

\begin{figure*}
  \centering
    \includegraphics[width=0.495\textwidth]{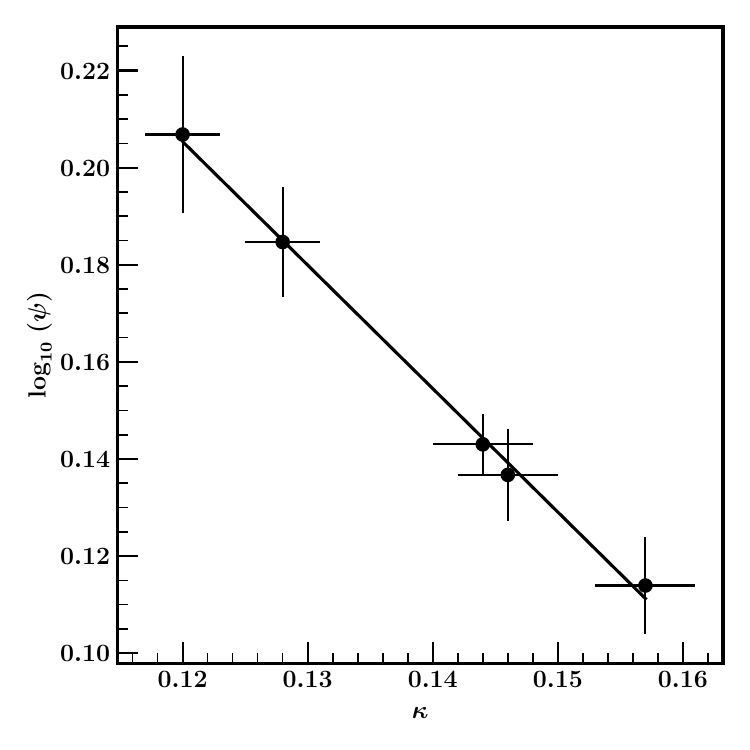}
    \includegraphics[width=0.495\textwidth]{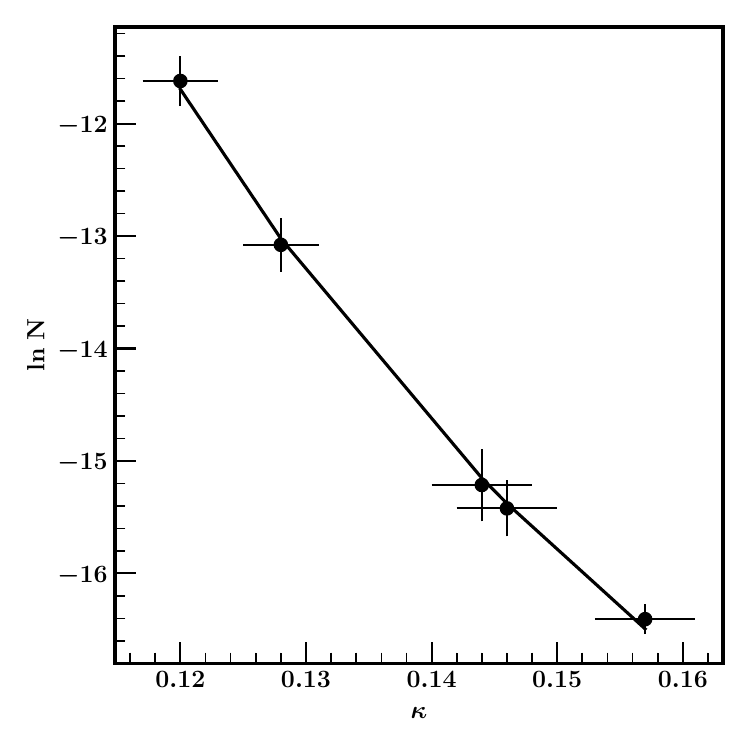}
      
   \caption{Scatter plots between the free parameters of the EDD model at five epochs. Left panel : $\kappa$ vs. $\log_{10} {\psi}$; right panel: $\ln{N}$ vs. $\kappa$. A solid curve in each panel is the best-fitted function as described in the text.}
   \label{fig:edd_cor_fit}
\end{figure*}

\autoref{fig:edd_cor_fit} shows the scatter plot between EDD model fit parameters. 
It is expected that $\log_{10}(\psi)$ will be inversely and linearly proportional to $\kappa$ since $\psi= \eta_R \xi_R^{-\kappa}$, and as expected we found a strong anti-correlation in $\psi$ vs  $\kappa$.

%We found a positive correlation in $\psi$ vs  Norm, $\psi$ vs.  B, and Norm vs. B, whereas a strong anti-correlation in $\psi$ vs  $\kappa$, $\kappa$ vs  Norm, and  $\kappa$ vs B. 
%We studied the correlations between free parameters of the EDD model, $\kappa$, $\psi$, and $\mathbb{N}$.
%The $\kappa$ and $\psi$ terms are related by $\psi= \eta_R \xi_R^{-\kappa}$ 
The above equation can be further expressed as 
\begin{equation} \label{eq:edd_psi_kappa}
    \log_{10}(\psi) = \log_{10}(\eta_R) - \kappa \times \log_{10}(\xi_R)
\end{equation}

\autoref{eq:edd_psi_kappa} was fitted using the values of $\kappa$ and $\psi$ obtained from the broadband SED fitting. The fitted line in the left panel of \autoref{fig:edd_cor_fit} on the $\kappa$ vs. $\log_{10}(\psi)$ plot is having a slope ($\log_{10}(\xi_R)$) of 2.54 and an intercept ($\log_{10}(\eta_R)$) of 0.51. This implies that the observed photon energy  $\xi_R^2 = 122$ MeV falls within the spectral coverage of our broadband spectra. 
We can estimate the $\gamma_R$ ($\xi_{R}$/$\sqrt{\mathbb{C}}$) and $\eta_R$ to be $3.6 \times 10^{8}$ and 3.24, respectively.

%We also estimate the $\gamma_R$ ($\xi_{R}$/$\sqrt{\mathbb{C}}$) and $\eta_R$ as $3.6 \times 10^{8}$ and 3.24, respectively.}
%--------------------------Old-----------------------
%\autoref{eq:edd_psi_kappa} was fitted using the values of $\kappa$ and $\psi$ obtained from the broadband SED fitting. The fitted line is shown in the black line in the left panel of \autoref{fig:edd_cor_fit} on the $\kappa$ versus $\log_{10}(\psi)$ plot, with the best-fitted line having a slope ($\log_{10}(\xi_R)$) of 2.598 and an intercept ($\log_{10}(\eta_R)$) of 0.549. This implies the observed photon energy corresponding to $\gamma_R$ is found to be $\xi_R^2 = 157.2$ MeV, which falls within the spectral coverage of our broadband spectra, and $\eta_R = 3.54$.
%-----------------------------------------------------------
%We fitted \autoref{eq:edd_psi_kappa} with the values of $\kappa$ and $\psi$ obtained from the broadband SED fitting. The fitted line is shown in the black line in the left panel of \autoref{fig:edd_cor_fit} on the $\kappa$ versus $\log_{10}(\psi)$ plot and the best-fitted line has slope ($\log_{10}(\xi_R)$) and intercept ($\log_{10}(\eta_R)$) as $2.598$ and 0.549, respectively. This implies the observed photon energy corresponding to $\gamma_R$ is found to be $\xi_R^2 = 157.2$ MeV, which is within the spectral coverage of our broadband spectra and $\eta_R = 3.54$.

\begin{figure*}
  \centering
    \includegraphics[width=0.495\textwidth]{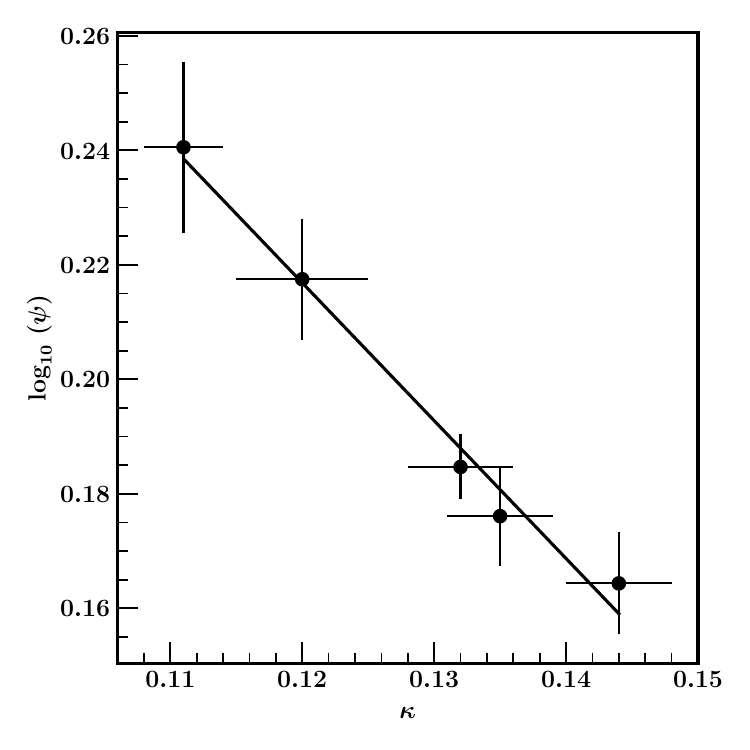}
    \includegraphics[width=0.495\textwidth]{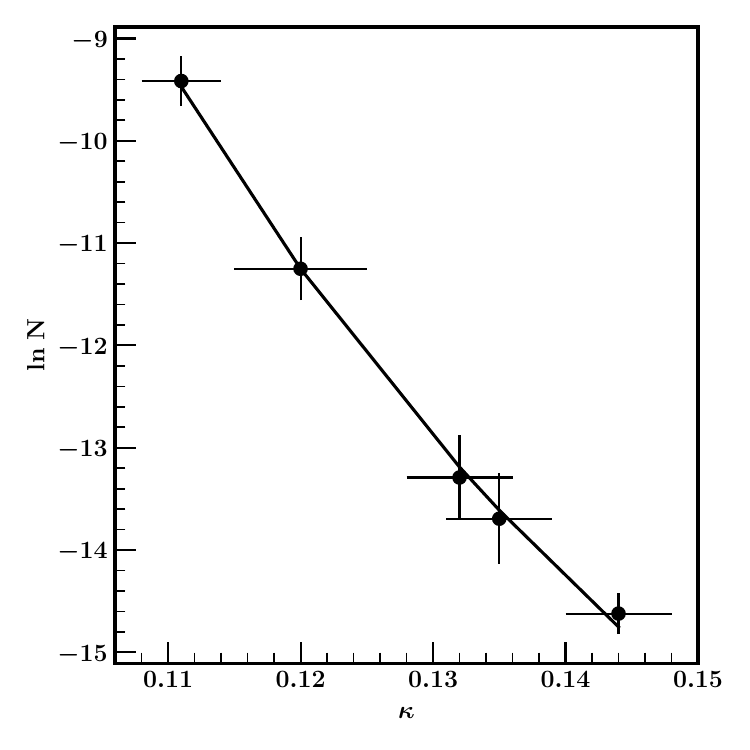}
      
   \caption{Scatter plots between the free parameters of the EDA model at five epochs. Left panel : $\kappa$ vs. $\log_{10} {\psi}$; right panel: $\ln{N}$ vs. $\kappa$. A solid curve in each panel is the best-fitted function as described in the text.}
   \label{fig:eda_cor_fit}
\end{figure*} 

%\begin{figure*}
  %\centering
 %   \includegraphics[width=0.495\textwidth]{figures/correlations/edd_gamma.pdf}
    %\includegraphics[width=0.495\textwidth]{figures/correlations/edd_kappa_N_revised.pdf}
      
  % \caption{Variation of $\gamma_o$ with bulk Lorentz factor ($\Gamma$) for the EDD model (left panel) at five epochs.}
   %\label{fig:edd_cor_fit}
%\end{figure*} 

Additionally, injection energy can be estimated using the correlation between the best-fit parameters, $\kappa$ and $\mathbb{N}$.
The relation between $\kappa$ and $\mathbb{N}$ is given in \autoref{eq:norm_edd} which can be further expressed as 
\begin{equation} \label{eq:edd_kappa_N}
    \ln(\mathbb{N}) = \frac{\eta_R}{\kappa}A^\kappa + B
\end{equation}

%\autoref{eq:edd_psi_kappa} 
\autoref{eq:edd_kappa_N} was fitted with the $\kappa$ and $\mathbb{N}$ values obtained from the broadband SED fitting. 

%\textbf{The fitted line is shown in the right panel of \autoref{fig:edd_cor_fit} on the $\kappa$ vs. $\ln (\mathbb{N})$ plot has $A = 9.66 \times 10^{-04}$, $B = -23.4$, and $\eta_R = 3.24$. Since $\gamma_0 \sim 9.66 \times 10^{-04} \gamma_R$, the estimated values suggest that the injection energy ($\gamma_0$ = $3 \times 10^{5}$) is significantly smaller than the $\gamma_R$. Such high value of minimum lorentz factor with standard one-zone SSC model for EHBLs sources is also reported by several authors \citep{2011Kaufmann,2021Zech,2024Goswami} .}

Using $\eta_R$ = 3.24 from above, we fitted the $\kappa$ vs. $\ln (\mathbb{N})$ plot shown in \autoref{fig:edd_cor_fit}, to obtain
 $A = 9.66 \times 10^{-04}$, $B = -23.4$. Since,  $\gamma_{0}$ $\sim$ $A\gamma_{R}$, we can estimate the  
injection energy, $\gamma_{0}$ $\sim$ $3.5 \times 10^{5}$, which is significantly smaller than $\gamma_{R}$. Such a high value of minimum Lorentz factor with a standard one-zone SSC model for EHBLs sources has also been reported by several authors \citep{2011Kaufmann,2021Zech,2024Goswami}.
%The injection energy $\gamma_0$ is turning out to be physically consistent.

%$\xi_{max}$=$\gamma_{max} \sqrt{\mathbb{C}}$
%--------------------------Old-----------------------
%\autoref{eq:edd_psi_kappa} was fitted with the $\kappa$ and $\mathbb{N}$ values obtained from the broadband SED fitting. The fitted line is shown in the black line in the right panel of \autoref{fig:edd_cor_fit} on the $\kappa$ versus $\ln (\mathbb{N})$ plot, and the best-fitted line has $A = 4.90 \times 10^{-08}$, $B = -17.34$, and $\eta_R = 3.54$. These values suggest that the injection energy $\gamma_0 \sim 4.90 \times 10^{-08} \gamma_R$ which is significantly smaller than the $\gamma_R$.
%-----------------------------------------------------------

%Here, we note that the best-fitted line is largely off from the best-fitted normalization value at one out of the five epochs of the observations. This offset might occur because of the quasi-simultaneous broadband spectra considered in this analysis.

%The fifth rows of \autoref{} include the Spearman's rank correlation results for the EDA model fit parameters, and 
\autoref{fig:eda_cor_fit} shows the scatter plot between the EDA model fit parameters. %A strong positive correlation is observed between index p and magnetic field with $r_s= 0.99$ ($P_{rs}= 0.00$), whereas a strong anti-correlation is noticed between index p and normalization and magnetic field B each with $r_s= -0.99$ ($P_{rs}= 0.00$). of the EDA model have a
The free parameters $\kappa$, $\psi$, and $\mathbb{N}$ in the EDA model are similar to the EDD model. 
%The free parameters in the EDA model are similar to the EDD model, $\kappa$, $\psi$, and $\mathbb{N}$.
The $\kappa$ and $\psi$ parameters of the EDA model follow a similar trend as in \autoref{eq:edd_psi_kappa}. However, the $\kappa$ and $\mathbb{N}$ parameters of the EDA model have a relation given in \autoref{eq:norm_eda} which can be further expressed as
\begin{equation} \label{eq:eda_kappa_N}
    \ln(\mathbb{N}) = \frac{\eta_R}{\kappa}A^\kappa - \kappa \ln(\xi_R) + B
\end{equation}

We fitted both \autoref{eq:edd_psi_kappa} and \autoref{eq:eda_kappa_N} with the $\kappa$, $\psi$, and $\mathbb{N}$ values of the EDA model. \autoref{eq:edd_kappa_N} was applied to the $\kappa$ vs. $\log_{10}(\psi)$ plot (left panel of \autoref{fig:eda_cor_fit}), while \autoref{eq:eda_kappa_N} was applied to the $\kappa$ vs. $\ln (\mathbb{N})$ (right panel of \autoref{fig:eda_cor_fit}).
From the fitted equations of the EDA model, we obtained $\xi_R^2 = 66 $ MeV, $\gamma_R =  2.66 \times 10^{8}$, $\eta_R = 3.21$, $A = 7.5 \times 10^{-04}$, and $B = -21.8$. With the relation, $\gamma_0 \sim 7.5 \times 10^{-04} \gamma_R$, we find $\gamma_0$ = $2 \times 10^{5}$. Thus the results obtained from the EDA model are qualitatively similar to those obtained for the EDD model.

\section{SUMMARY AND DISCUSSION} \label{sec:summary}

In this work, we have performed a detailed broadband SED analysis of an EHBL source, 1ES 0229$+$200, using simultaneous multi-wavelength observation taken at different epochs from September 2017 to August 2021 (MJD 58119$-$59365) using {\em AstroSat}$-$LAXPC, SXT, and UVIT. We have also included the $\gamma$-ray data from {\em Fermi}-LAT observed from August 2008 to October 2022 (MJD 54682.6$-$59882) and VHE $\gamma$-rays data of MAGIC observed from 2013 to 2017 (MJD 56293 $-$ 58118). We used the one-zone synchrotron and SSC model (\emph{$sscicon \otimes n(\xi)$}) with various particle distributions viz. log parabola, broken power law, power law with maximum gamma ($\gamma_{max} $),  energy-dependent diffusion (EDD) and energy-dependent acceleration (EDA) model to fit the broadband SED. 

%This study suggests that the SED of 1ES 0229$+$200 for all epochs with all particle distributions can be well explained by the one-zone leptonic model up to the GeV energy range, but we obtained a factor of two changes in the case of VHE MAGIC spectra. This might occur because the VHE MAGIC data is quasi-simultaneous with other observations.

%Considering a one-zone SSC model, the data exhibit a good fit because the source is BL Lac.
%The jet parameters obtained at different epochs are consistent with each other and with previous studies, where SED modelling for this source was performed. 
%However, according to  \citet{2018Costamante,2021Zech}, the broadband SED modelling in hard-TeV blazars can be explained by the one-zone SSC model at the cost of extremely high electron energies and a significantly low magnetization value with very low radiative efficiency. They estimated an average $\gamma_{max}$ as $\sim10^7$ for six hard-TeV blazars. However, we found a relatively higher value of $\gamma_{max}$ ($10^8$) and consistent magnetic field strength (in the range of a few mG) for 1ES 0229$+$200.

According to \citet{2018Costamante}, the broadband SED modelling in hard-TeV blazars can be explained by the one-zone SSC model with a smooth broken power-law particle distribution.
%They estimated the average maximum energy of the electrons, $\gamma_{max}$ as  $\sim10^7$, break energy of the electrons, $\gamma_{break}$ as  $\sim10^6$ and magnetic field strengths of in the range of a few mG for six hard-TeV blazars. However, for TeV blazar 1ES 0229$+$200, assuming a relatively higher value of $\gamma_{max}$ ($10^8$) we found magnetic field strength in the range of a few mG, and break energy of the electrons in the order of $\sim10^6$.}
They estimated the break energy of the electrons, $\gamma_{break}\sim10^6$ and magnetic field strengths in the range of a few mG for six hard-TeV blazars.  In this work, for the TeV blazar 1ES 0229$+$200, we also find the magnetic field strength in the range of a few mG and the break energy of the electrons in the order of $\sim10^6$.

%We calculated the the Jet power, $P_{jet}$ for different particle energy distributions at the five epochs. In case of broken power-law and $\xi_{max}$ models, we found $P_{jet}$ as $\sim$10$^{47}$ erg s$^{-1}$ with a minimum Lorentz factor of $\gamma_\text{min} = 10$ and it reduces to $\sim 10^{44}$ erg s$^{-1}$ when increasing $\gamma_\text{min}$ to $10^4$. However, the $P_{jet}$ for the other energy particle distributions with intrinsic curvature was found to be $\sim 10^{44}$ erg s$^{-1}$, which is independent of $\gamma_\text{min}$. We found that the estimated $P_{jet}$ is almost independent of the bulk Lorentz factor $\Gamma$ and size R. For the intrinsically curved particle energy distributions (e.g., log parabola, EDD and EDA models), the estimated $P_{jet}$ ($\sim$10$^{44}$ erg s$^{-1}$) is only a small fraction of the Eddington luminosity (1.26 $\times$ $10^{47}$ erg s$^{-1}$) of the blazar's black hole mass \citep[$10^9$ $M_\odot$; ][]{2012Meyer}, suggesting that the accretion processes may power the jet. The above estimated $P_{jet}$ value is in good agreement with the previous estimate by \citet{2020Acciari} for the SSC model of the source. \textbf{Similar results have been reported recently by \citet[][under review]{2024hritwik}, where they calculated the jet power for the HBL source Mrk 501 by considering the same particle distribution and found the jet power is $\sim$10$^{42}$ erg s$^{-1}$}.

We computed the jet power, $P_{\text{jet}}$, for various particle energy distributions at five different epochs. In the case of broken power-law and PL with $\gamma_{\text{max}}$ models, we find $P_{\text{jet}} \sim 10^{47}$ ergs/sec  with a minimum Lorentz factor ($\gamma_{\text{min}}$) set to 10. This value decreases to approximately $10^{44}$ ergs/sec  when $\gamma_{\text{min}}$  is increased to $10^4$. However, for other particle energy distributions with intrinsic curvature, the calculated $P_{\text{jet}}$ remains around $10^{44}$ ergs/sec, irrespective of $\gamma_{\text{min}}$. Interestingly, we found that the estimated $P_{\text{jet}}$  is nearly independent of the bulk Lorentz factor ($\Gamma$) and size (R). In the case of intrinsically curved particle energy distributions, such as the log parabola, EDD, and EDA models, the $P_{\text{jet}}$ ($\approx$$10^{44}$ ergs/sec)  represents only a small fraction of the Eddington luminosity (1.26 $\times 10^{47}$  ergs/sec) of the blazar's black hole mass \citep[$10^9$ $M_\odot$; ][]{2012Meyer}, suggesting that accretion processes might be driving the jet.
 
For the source 1ES 0229$+$200, \citet{2020Acciari} estimated $P_{\text{jet}}$ ($\approx$$10^{44}$ ergs/sec) applying the one zone SSC model assuming a power law with exponential cutoff particle distribution which agrees well with our estimation.
%This $P_{\text{jet}}$ value agrees well with previous estimates by \citet{2020Acciari} of the source for the one zone SSC model with a power law with exponential cutoff particle distribution.
Consistent findings have been reported recently by \citet{2024Bora}. They calculated the jet power for the HBL source Mrk 501 using the same particle distributions. %According to \citet{2024Bora} 
They found that the estimated jet power with a broken power-law distribution was around $10^{47}$ ($10^{44}$) ergs/sec with a minimum electron energy of $\gamma_{min}$ = 10 ($10^3$). However, the estimated jet power was found to be considerably lower than a few times ($10^{42}$ ergs/sec) for electron energy distributions with intrinsic curvature (such as the log-parabola form).

Correlation studies among the best-fit model parameters provide significant insights regarding the consistency and adequacy of the model in describing the observed broadband SED. For the power-law with a maximum gamma ($\gamma_{max}$ ) model, we find a strong anti-correlation between the index p and normalisation $\mathbb{N}$ and a positive correlation between index p and magnetic field B. As discussed in \citet{2021HOTA}, these correlations are against expectations. This is because, in this model, the change in normalization (\autoref{eq:norm_ximax}), $\Delta \mathbb{N}/\mathbb{N} = log (\xi_0^2) \Delta p /2$, where $\xi_o^2 \equiv \gamma_0^2\mathbb{C}$, occurs for a change in the index, $\Delta p$. Since $\gamma_o$ is much smaller than the $\gamma$ required to produce X-ray photons, $ log (\xi_0^2)$ should be large; the normalization should vary significantly with a positive correlation when the index changes. However, we find that the normalisation varies with the change in the index with a strong anti-correlation. The observed anti-correlation could be explained by a more sophisticated model in which the acceleration time scale and magnetic field are associated.

%Since the log-parabola model is a simpler version of a more extensive physical model, we consider physical models that may reproduce the curvature. We investigated the scenario of PL with $\xi_{max}$ model and observed a strong anti-correlation between the index p and normalisation N that conflicts with the model prediction by functional form of normalisation. Further, we also observed a strong positive correlation between index p and magnetic field B, which contradicts the model prediction. The variation in the model parameters, p and B, depends on the variation of the acceleration time scale, suggesting an anti-correlation between p and B. %Such a positive correlation can occur if it is postulated that the acceleration time scale varies directly with the magnetic field.

%\textbf{We investigated the scenario of PL with the maximum electron energy model and observed a strong anti-correlation between the index (p) of the electron distribution and normalisation that conflict with the model prediction by functional form of normalisation. Further, a strong positive correlation is noticed between the index p and the magnetic field B, which also contradicts the model prediction where the variation in the observed parameters depends on the variation of the acceleration time-scale, which in turn suggests an anti-correlation between p and B. Such a positive correlation can occur if it is postulated that the acceleration time scale varies inversely with the magnetic field.}

We showed that the spectral curvature may also be reproduced by the energy-dependent electron diffusion (EDD) model, where we consider the escape or diffusion time scale to be energy-dependent. As expected by the predictions of the model, we found a strong anti-correlation between the two model parameter normalization and $\psi$ with the third parameter $\kappa$. The compatibility of the mentioned correlation with the predicted one, for $\Gamma = 20$, allows us to calculate the reference photon energy $\xi_R^2$ = 122 MeV, arising from an electron with energy $\gamma_R =  3.6 \times 10^{8}$, and the energy of the electrons that are injected into the acceleration region, as $\gamma_0$ =  $3.5 \times 10^{5}$. Similarly, we consider the spectral curvature may also be reproduced by the energy-dependent electron acceleration (EDA) model, where the acceleration time scale is energy-dependent. For the EDA model, the calculated values are approximate as $\xi_R^2= 66$ MeV, $\gamma_R =  2.66 \times 10^{8}$, and $\gamma_0$ =  $2 \times 10^{5}$. 

In a previous investigation, \citet{2021HOTA} utilized the EDD and EDA models to analyze the X-ray observations of the HBL source, Mkn 421, during its flaring phase, deriving respective spectral parameters. They identified a different range of spectral parameters, with $\kappa$ falling between 0.3 and 1.0, and $\psi$ between 1.12 and 1.96. These findings resulted in a slope of $\log_{10} \xi_{R}$ = 0.38 for the $\kappa$ vs. $\log \psi$ anti-correlation. However, in our analysis of EHBL source 1ES 0229+200 (present study), we observe $\kappa$ and $\psi$ values within the ranges of 0.11$-$0.16 and 1.30$-$1.65, respectively. The correlation between these parameters yields a steeper slope, $\log_{10} \xi_R$ = 2.54. Consequently, we obtain a significantly higher value of $\xi_R^2$ ($10^4$ times larger than the HBL source). Notably, the calculated values of $\xi_R^2$ in both the EDD and EDA models predominantly fall within the MeV range, surpassing the Synchrotron spectral energy range under consideration. This suggests favourable conditions for these models to fit the broadband spectrum of EHBL sources effectively. In contrast, for Mkn 421 during its flaring phase, the corresponding photon energy ($\xi_R^2$ $\sim$ 5.6 keV) remains within the considered energy range (0.5 $-$18 keV) in both the EDD and EDA models, contradicting the spectral fitting of HBL sources with these models. Additionally, we get a relationship $\gamma_{0}$ = 9.6 $\times$ $10^{-04}$ $\gamma_{R}$ when fitting the correlation between $\kappa$ and ln $\mathbb{N}$, which is more likely appropriate in the physical framework than it is observed for the HBL source (i.e $\gamma_{0}$ $\sim$ 0.2 $\gamma_{R}$) \citep{2021HOTA}. Moreover, we can now get the actual values of $\gamma_R$ and, more importantly, $\gamma_0$ while using the SSC model (current study).

The intrinsically curved particle distribution models (EDD and EDA models) considered in this work are simple and have analytical solutions; the physical situation may really be more complicated. Furthermore, considering both escape and acceleration time scales to be energy-dependent would be more physically plausible. Compared to the power-law form used in this work, the energy dependency of various time scales may differ. Eventually, in order to provide a more complete picture, the analysis must be expanded to include some other EHBL blazars. Moving forward, it would be valuable to explore additional aspects, such as temporal variations and extended datasets, to further refine our understanding of the jet dynamics and uncover any nuanced behaviours that might emerge over an extended observation period. Additionally, comparative analyses with other blazar sources could provide a comprehensive perspective on the universality of the observed trends in jet power and its dependence on particle energy distributions.

\section*{Acknowledgements}
We thank the anonymous referee for insightful comments and constructive suggestions. The authors JH and ACP would like to acknowledge Inter-University Centre for Astronomy and Astrophysics (IUCAA), Pune, India, for providing facilities to carry out this work. This publication uses data from the {\em{Astrosat}} mission of the Indian Space Research Organisation(ISRO), archived at the Indian Space Science Data Centre(ISSDC). This work has used the data from the Soft X-ray Telescope (SXT) developed at TIFR, Mumbai. LaxpcSoft software is used for analysis of the LAXPC data and we acknowledge the LAXPC Payload Operation Center (TIFR, Mumbai). This research has made use of data, software and/or web tools obtained from the High Energy Astrophysics Science Archive Research Center (HEASARC), a service of the Astrophysics Science Division at NASA/GSFC and of the Smithsonian Astrophysical Observatory’s High Energy Astrophysics Division.

%%%%%%%%%%%%%%%%%%%%%%%%%%%%%%%%%%%%%%%%%%%%%%%%%%
\vspace{5mm}
\facilities{ Fermi(LAT), AstroSat (UVIT, SXT AND LAXPC), Swift(UVOT)}


\software{ fermipy$-$v0.17.4 (\url{Fermipy webpage: https://fermipy.readthedocs.io/en/latest/}),\\
%{\em{AstroSat}} Science Support Cell\url{http://astrosat-ssc.iucaa.in},\\ 
XSPEC (\url{https://heasarc.gsfc.nasa.gov/xanadu/xspec/}),\\ HEASARC(\url{https://heasarc.gsfc.nasa.gov/docs/software/heasoft/})
          }



\bibliographystyle{aasjournal}
\bibliography{example} % if your bibtex file is called example.bib

%\bibliography{sample631}{}
%\bibliographystyle{aasjournal}



\end{document}